%% file: main.tex
\documentclass[prx,twocolumn,english,superscriptaddress,floatfix,longbibliography]{revtex4-2}
\usepackage{amsfonts}
\usepackage{graphicx}
\usepackage{dcolumn}
\usepackage{bm}
\usepackage{enumitem} 
\usepackage{amsthm} 
\usepackage{subcaption} 
\usepackage{mathtools} 
\usepackage{rotating} 
\usepackage{thmtools, thm-restate} 
\usepackage{comment}

\usepackage[normalem]{ulem} 

\usepackage{tikz}
\usetikzlibrary{arrows,
                chains,
                positioning,
                shapes}
                
\tikzstyle{io} = [trapezium, trapezium left angle=70, trapezium right angle=110, minimum width=1cm, minimum height=1cm, text centered, draw=black, fill=blue!20]
\tikzstyle{decision} = [diamond, minimum width=3cm, minimum height=1cm, text centered, text width = 2cm, inner sep = 0pt, draw=black, fill=blue!20]
\tikzstyle{block} = [rectangle, draw, fill=blue!20, 
    text width=5cm, text centered, rounded corners, minimum height=4cm]
\tikzstyle{process} = [rectangle, minimum width=3cm, minimum height=1cm, text centered, text width=4.75cm, draw=black, fill=blue!20]
\tikzstyle{processDiag} = [rectangle, minimum width=3cm, minimum height=1cm, text centered, text width=4.75cm, draw=black, fill=yellow!60]
\tikzstyle{arrow} = [thick,->,>=stealth]
\tikzstyle{arrowDiag} = [thick,->,brown,>=stealth ]
\tikzstyle{cloud} = [draw, ellipse,fill=red!20, node distance=3cm,
    minimum height=2em]

\theoremstyle{definition}
\newtheorem{definition}{Definition}
\newtheorem{theorem}{Theorem}
\newtheorem{corollary}{Corollary}[theorem]
\newtheorem{lemma}[theorem]{Lemma}

\newcommand{\ket}[1]{|#1\rangle}             
\newcommand{\bra}[1]{\langle#1|}             

\newcommand{\norm}[1]{\left \vert\!\left \vert #1 \right \vert\!\right\vert}             
\newcommand{\abs}[1]{\left \vert \! #1 \right \vert}             
\newcommand{\Tr}[1]{\textrm{Tr}\left [ #1 \right ]}             
\newcommand{\tr}[2]{\textrm{Tr}_{#2}\left [ #1 \right ]}             

\newcommand{\eyeState}{\mathbb{I}_\mathcal{H}} 
\newcommand{\eyeMeas}{\mathbb{I}_\mathcal{K}} 

\newcommand{\rhomodel}{\rho_{\mathrm{model}}}
\newcommand{\rhoPR}{\rho_{\mathrm{PR}}}
\newcommand{\rhophi}[1]{\rho_{\phi_{#1}}}
\newcommand{\tauphi}[1]{\tau_{\phi_{#1}}}

\newcommand{\intd}[1]{\int^{2\pi}_0 \mathrm{d}\phi_{#1}\ }
\newcommand{\intdn}{\int^{2\pi}_0 \ldots \int^{2\pi}_0 \mathrm{d}\phi_{1}\ldots\mathrm{d}\phi_{n}\ }

\newcommand{\tp}[1]{\tilde{p}_{\Phi_{#1}}(\phi_{#1}\vert \phi_1 \ldots \phi_{#1-1})}
\newcommand{\tpo}{\tilde{p}_{\Phi_1}(\phi_{1})}
\newcommand{\tpn}{\tilde{p}_{\Phi_1\ldots\Phi_n}(\phi_{1}\ldots \phi_{n})}
\newcommand{\p}[1]{p_{\Phi_{#1}}(\phi_{#1}\vert \phi_1 \ldots \phi_{#1-1})}
\newcommand{\po}{p_{\Phi_1}(\phi_{1})}
\newcommand{\pn}{p_{\Phi_1\ldots\Phi_n}(\phi_{1}\ldots \phi_{n})}

\newcommand{\Uphi}[1]{U_{\phi_{#1}}}
\newcommand{\Udagphi}[1]{U_{\phi_{#1}}^\dag}

\newcommand{\deltaleak}{\delta_{\text{leak}}}

\begin{document}


\title{Imperfect Phase-Randomisation and Generalised Decoy-State Quantum Key Distribution}

\author{Shlok Nahar}
\email{sanahar@uwaterloo.ca}
 \affiliation{Institute for Quantum Computing and Department of Physics and Astronomy, University of Waterloo,
Waterloo, Ontario N2L 3G1, Canada}
\author{Twesh Upadhyaya}
\affiliation{Institute for Quantum Computing and Department of Physics and Astronomy, University of Waterloo,
Waterloo, Ontario N2L 3G1, Canada}
\author{Norbert L\"utkenhaus}
\affiliation{Institute for Quantum Computing and Department of Physics and Astronomy, University of Waterloo,
Waterloo, Ontario N2L 3G1, Canada}

\date{\today}

\begin{abstract}

Decoy-state methods \cite{hwang2003quantum, lo2005decoy, wang2005beating} are essential to perform quantum key distribution (QKD) at large distances in the absence of single photon sources. However, the standard techniques apply only if laser pulses are used that are independent and identically distributed (iid). Moreover, they require that the laser pulses are fully phase-randomised. However, realistic high-speed QKD setups do not meet these stringent requirements \cite{grunenfelder2020performance}.
In this work, we generalise decoy-state analysis to accommodate laser sources that emit imperfectly phase-randomised states. We also develop theoretical tools to prove the security of protocols with lasers that emit pulses that are independent, but not identically distributed. These tools can be used with recent work \cite{curty2022phasedecoy} to prove the security of laser sources with correlated phase distributions as well. We quantitatively demonstrate the effect of imperfect phase-randomisation on key rates by computing the key rates for a simple implementation of the three-state protocol.

\end{abstract}

\maketitle


\section{\label{sec:intro}Introduction}

Quantum key distribution (QKD) is a method to realise quantum-safe cryptography \cite{mosca2018cybersecurity}. Since QKD does not rely on computational assumptions, QKD protocols can be proved to be information-theoretically secure \cite{mayers1996quantum, mayers2001unconditional, renner2005information}. However, practical implementations suffer from security loopholes which arise from a gap between the theoretic models for which security is proved, and the experimental devices that perform QKD \cite{lucamarini2018implementation}. Thus, better modelling of devices as well as theoretical tools to perform security analysis with these more detailed models is essential for the implementation security of QKD protocols.

There have been recent advances to theoretically accommodate general source imperfections \cite{pereira2019quantum,pereira2020quantum,navarrete2021practical}. However, these techniques cannot be used at present with the decoy-state method \cite{hwang2003quantum, lo2005decoy, wang2005beating} which is essential to get secret key rates at large distances with coherent states. To this end, there has been more work on doing decoy-state QKD with intensity correlations \cite{zapatero2021security, sixto2022security}.

Besides the absence of intensity correlations, standard decoy-state methods still assume that the laser outputs fully phase-randomised states. For phase-randomisation in gain-switched laser diodes, it is essential that no photons from previous pulses remain in the lasing cavity at the start of the next lasing \cite{kobayashi2014evaluation}. Such an assumption has been demonstrated to not hold for lasers with a high repetition rate \cite{grunenfelder2020performance} as there is not enough time for the laser cavity to empty out between pulses.
In this case, the laser pulses might even have correlated phases distributions.

Techniques to prove the security of decoy-state QKD in the presence of phase correlations were developed in \cite{curty2022phasedecoy}. They use a novel proof technique that reduces the security analysis of phase-correlated laser pulses to that of laser pulses that have an independent and non-identically distributed phase distribution.
The decoy-state analysis for such phase-independent states was first described by one of the authors in \cite{nahar2022decoy}.
Note that as shown in \cite{curty2022phasedecoy} the proof techniques described in \cite{nahar2022decoy} only work for laser pulses that have no phase correlations.

In Section \ref{sectionSourceMapCorrLaser}, we develop tools to reduce the security of phase-independent laser pulses to the security of an iid partially phase-randomised laser. For this reduction, the phase distribution of the laser pulses must be partially characterised by a single parameter. This is practically useful only if this parameter is experimentally measurable. So, we include a discussion on the methods and difficulties of measuring this quantity in Section \ref{sectionCharLaser}.

In order to use decoy-state analysis for partially phase-randomised laser pulses, we draw an analogy with channel tomography to state our generalised decoy-state methods in Section \ref{sectionGenDecoy}. However, as outlined with the requisite background in Section \ref{secBackground}, a full decoy-state analysis in this approach requires the diagonalisation of the laser states. Thus, we describe how to approximately diagonalise a density matrix in Section \ref{approxDiag}.

In summary, we develop tools that enable us to perform decoy-state QKD with lasers that have imperfect phase-randomisation. We then use these tools to analyse the security of the three-state protocol with phase imperfections in Section \ref{sec3StateProtocol} as an example. We also plot our results to depict the effect of the phase imperfections for this protocol.


\section{Background} \label{secBackground}

In this section we first summarise the steps in a generic prepare and measure (PM) protocol. We then review the key rate optimisation problem, and discuss the concept of source maps, a proof technique used to find lower bounds on the key rate.

\subsection{QKD protocol steps}

Here we outline the steps in a generic PM protocol with $n$ rounds. We focus in particular on the asymptotic limit, where the number of protocol rounds $n$ tends to infinity.
\begin{enumerate}
    \item \label{prep} \textbf{State Preparation:} Alice randomly prepares one of a set of quantum states $\{\rho^\mu_1 \ldots \rho^\mu_n\}$ with an a priori probability distribution $\{p(i,\mu)\}$, where the signal modulation $i$ and signal intensity $\mu$ denote which state she chose. The prepared states are called signal (and decoy) states. We model Alice's signal preparation procedure as different channels acting on some fixed base state $\rho^\mu_i = \Xi_i (\rho^\mu)$. We denote the quantum system associated with each of these signal states $A'_m$ for the $m^\textrm{th}$ round of the protocol.
    \item \textbf{Signal Transmission:} Alice sends her prepared states to Bob via an insecure quantum channel
    \begin{align*}
        \mathcal{E} : A'_1\ldots A'_n \xrightarrow[]{} B_1\ldots B_n
    \end{align*}
    where each $B_m$ denotes the quantum system associated with each of the states Bob receives in round $m$.
    \item \label{meas} \textbf{Measurement:} Bob measures the states that he receives by a $k$-outcome POVM $\{\Gamma_j\}_{j=1}^k$ and records the outcome from each round.
\end{enumerate}
After repeating the above steps multiple times, we proceed to the next part of the protocol.
\begin{enumerate}[resume]
    \item \label{paramEst} \textbf{Acceptance testing:} Alice and Bob randomly choose a subset of the rounds for testing. For the rounds chosen for testing, they both publicly announce the signal modulation $i$ and signal intensity $\mu$ chosen, and measurement outcome $j$ to form a frequency distribution. They then check if this frequency distribution belongs to the acceptance set agreed upon before running the protocol. If it does, they proceed with the protocol after discarding the test results. Otherwise, they abort.
    
    In the asymptotic limit, assuming that Eve's attack is iid, i.e. $\mathcal{E} = \Phi^{\otimes n}$, the frequency distribution converges to a probability distribution $p(i,\mu,j)=p(i,\mu)\Tr{\Gamma_j\Phi(\rho^\mu_{i})}$.
    This probability distribution effectively constrains $\Phi$, and thus Eve's actions on the states that Alice sent Bob. 
    \item \label{announce} \textbf{Announcements and sifting:} Alice and Bob make announcements over the authenticated classical channel. They sift the non-tested data based on the announcements made i.e. they choose a subset of signal and measurement data to keep and discard the rest based on the announcements.
    \item \label{keyMap} \textbf{Key Map:} Alice uses her signal modulation data $i$ as well as the announcements to map her data into a key string $x$. This is called the raw key. We assume here that the key is a bit string for simplicity, but all the steps can be applied more generally.
    \item \label{protocolStepErrorCorr} \textbf{Error Correction:} Alice and Bob then perform error correction over the authenticated classical channel to make Bob's measurement outcomes match with Alice's bit string $x$. We denote the data communicated per key bit to Eve in this process as $\delta_\textrm{leak}$.
    \item \textbf{Privacy Amplification:} Alice and Bob produce their final secret key by applying an appropriate hash function on the raw key (Theorem 5.5.1 of \cite{renner2008security}).
\end{enumerate}
PM protocols are typically implemented in experiments. However, it is easier to analyse the security of another class of protocols, entanglement-based (EB) protocols where Alice and Bob share an entangled bipartite state instead of step \ref{prep} of the PM protocol.

Fortunately, we can reduce the analysis of any PM protocol to the analysis of an EB protocol with added constraints via a source-replacement scheme \cite{bennett1992quantum,curty2004entanglement,li2020improving} as follows.
First, define $\ket{\rho^\mu_i}_{A_SA'}$ to be a purification of $\rho^\mu_i$. The purifying system $A_S$, termed the shield system \cite{horodecki2009general}, is useful for the security proof if Alice sends Bob mixed states. Neither Alice nor Bob interacts with the shield system at any point.

Alice prepares the state
\begin{align} \label{eqPsiAAsA'}
    \ket{\psi}_{AA_SA'} = \sum_{i,\mu} \sqrt{p(i,\mu)} \ket{i,\mu}_A \otimes \ket{\rho^\mu_i}_{A_S A'}
\end{align}
and sends system $A'$ to Bob through the insecure quantum channel to get the state $\rho_{AA_SB}$.
In addition to the constraints from step \ref{paramEst} of the protocol that take the form $p(i,\mu,j) = \Tr{\left(\ket{i,\mu}\bra{i,\mu}_A\otimes\mathbb{I}_{A_S}\otimes\Gamma_j\right)\  \rho_{AA_SB}}$, we get the constraint $\tr{\rho_{AA_SB}}{B} = \tr{\ket{\psi}\bra{\psi}_{AA_SA'}}{A'}$. Intuitively, this represents the fact that Eve cannot change the states in Alice's lab and shield system, although she can act freely on the state sent to Bob. We shall now briefly outline how we can use these constraints to reliably lower bound the secret key rate that we can obtain from a QKD protocol.

\subsection{Numerical Asymptotic Key Rate} \label{secNumKeyRate}

The secret key rate in the asymptotic limit under the iid assumption can be found using the Devetak-Winter formula: $R^\infty = H(Z\vert E)-\deltaleak$ where $Z$ is the key register, $E$ is Eve's register, and $\deltaleak$ is the number of bits per round leaked to Eve during Step \ref{protocolStepErrorCorr} of the protocol. A lower bound for the key rate can be found by minimising the first term over all possible marginal states that Eve could hold. The Devetak-Winter key rate can be lifted to coherent attacks if the protocol is permutation-invariant via the quantum de Finetti theorem \cite{renner2007symmetry} or the postselection technique \cite{christandl2009postselection}.

Following \cite{coles2016numerical,winick2018reliable} the Devetak-Winter key rate formula for an EB protocol can be reformulated as an SDP
\begin{equation}
    \begin{aligned}
        R^\infty = \underset{\rho_{AA_SB}}{\textrm{min}} &D\left(\mathcal{G}\left(\rho_{AA_SB}\right)\vert\vert\mathcal{Z}\left(\mathcal{G}\left(\rho_{AA_SB}\right)\right)\right) - \delta_\textrm{leak} \\
        \textrm{s.t. } &\textrm{Tr}\left[\Gamma_j\Phi(\rho^\mu_i)\right] = \gamma_{j\vert i,\mu} \quad\quad\forall i,j,\mu\\
        & \textrm{Tr}_B\left[\rho_{AA_SB}\right] = \rho_{AA_S}
    \end{aligned}
\end{equation}
where $A$, $A_S$ and $B$ are Alice and Bob's registers together with the shield system. The statistics $\gamma_{j\vert i,\mu}$ can be understood to be the conditional probability of Bob observing outcome $j$ given that Alice sent signal state $i$ and intensity $\mu$. Here, the relative entropy $D\left(\mathcal{G}\left(\rho_{AA_SB}\right)\vert\vert\mathcal{Z}\left(\mathcal{G}\left(\rho_{AA_SB}\right)\right)\right)$ is the objective function where $\mathcal{G}$ is a map that represents the protocol (including announcements), and $\mathcal{Z}$ is a map that can be constructed from the key map.

Since we do not use most of the specific details of these maps, we abstract the objective function as $f(\rho_{AA_SB})$. For details, see \cite{winick2018reliable, coles2016numerical}.
Note that although $\rho_{AA_SB}$ contains all signal and decoy intensities $\mu$, we choose to include only the signal intensity $\mu=\mu_S$ in the objective function for computational simplicity by using a key map that assigns key value only for the signal intensity $\mu_S$.

This SDP is infinite-dimensional, and so following Eq. (49) from \cite{upadhyaya2021dimension} we use the dimension reduction method. This technique involves taking a projection $\Pi_N$ onto the subspace containing less than $N+1$ photons, to construct a finite-dimensional SDP that would lower bound the infinite-dimensional SDP. The finite-dimensional SDP is given as
\begin{equation}
    \begin{aligned}
        R^N = \underset{\rho_{AA_SB}^N}{\textrm{min}} &f(\rho_{AA_SB}^N)-\delta_\textrm{leak} \\
        \textrm{s.t. }&\gamma_{j\vert i,\mu}-W^\mu\leq\textrm{Tr}\left[\Gamma_j^N\Phi(\rho^\mu_i)\right] \leq \gamma_{j\vert i,\mu} &\forall i,j,\mu\\
        & 1-W\leq \Tr{\rho_{AA_SB}^N}\leq 1\\
        & \textrm{Tr}_B\left[\rho_{AA_SB}^N\right] \leq \rho_{AA_S}
    \end{aligned}
\end{equation}
where $\rho_{AA_SB}^N = (\mathbb{I}_{AA_S}\otimes\Pi_N)\rho_{AA_SB}(\mathbb{I}_{AA_S}\otimes\Pi_N)$ and $\Gamma_j^N = \Pi_N\Gamma_j\Pi_N$. $W$ is a parameter that needs to be estimated from Bob's observations that signifies the weight of $\rho_{AA_SB}$ that lies outside the subspace we are projecting on i.e. $W \geq 1-\Tr{\rho_{AA_SB}^N}$.
Note that we have used $\left[\Gamma_j,\Pi_N\right] = 0$ to obtain tighter constraints. This condition is commonly satisfied when we talk about photon-counting receiver modules that are block-diagonal in the total photon number. However, it is not crucial to use this and more details on obtaining the key rate for the fully general case are given in \cite{upadhyaya2021dimension}.

The SDP can be further simplified if the signal states have some block-diagonal structure and can be written as a direct sum $\rho^{\mu_S}_i= \bigoplus_{\tilde{n}=0}^\infty p_{\tilde{n}} \rho_i^{\tilde{n}}$ where the block-diagonal structure is the same for all the signals $i$. Here, $\mu_S$ denotes the signal intensity. This is obviously the case when we use fully phase-randomised states where $\ket{\tilde{n}}$ directly represents the photon number. As we shall show in Section \ref{secApptoDecoy}, we can obtain similar structure with partially phase-randomised states as well.

Following Eq. (D.6) and Eq. (D.9) from \cite{li2020application}, we can exploit the block-diagonal structure to write $f(\rho_{AA_SB}) = \sum_{\tilde{n}=0}^\infty p_{\tilde{n}} f(\rho_{AB}^{\tilde{n}})$ as a sum of positive terms. Thus, taking finitely many of these terms is sufficient to lower bound the key rate. In practice, just one of these terms is usually enough to give a good bound on the key rate for most protocols. For example, in a standard decoy-state protocol with fully phase-randomised states, considering just the term corresponding to single photons is sufficient to give a useful lower bound on the key rate.

If we could find the statistics to constrain each of these terms as $Y^L_{\tilde{n}}(i,j)\leq\Tr{\Gamma_j^N\Phi(\rho^{\tilde{n}}_i)}\leq Y^U_{\tilde{n}}(i,j)$, we could obtain the set of SDPs
\begin{equation}
    \begin{aligned} \label{blockKeyRate}
        R_{\tilde{n}}^N = \underset{\rho^{\tilde{n} N}_{AB}}{\min}\;  & p_{\tilde{n}} f(\rho^{\tilde{n} N}_{AB})\\
        \textrm{s.t. } &  Y_{\tilde{n}}^L(i,j)\leq \Tr{\Gamma_j^N\Phi(\rho^{\tilde{n}}_i)}\leq Y_{\tilde{n}}^U(i,j)\\
        & \tr{\rho_{AB}^{\tilde{n}N}}{B} \leq \rho^{\tilde{n}}_{A}\\
        & 1-W_{\tilde{n}} \leq \Tr{\rho_{AB}^{\tilde{n}N}}\leq 1\\
        & \rho_{AB}^{\tilde{n} N}\geq 0
    \end{aligned}    
\end{equation}
which can be related to the key rate as $R^N = \sum_{\tilde{n}} R_{\tilde{n}}^N - \delta_{\textrm{leak}}.$ Note that solving each of these SDPs independently will introduce some looseness since we do not take into account the fact that the constraints of different blocks are in general correlated. We describe new methods to upper and lower bound $\Tr{\Gamma_j^N\Phi(\rho^{\tilde{n}}_i)}$ via the generalised decoy-state analysis described in Section \ref{sectionGenDecoy}.

To summarise, if we have an iid protocol, signal states that are all block-diagonal in the same basis, and we have bounds on the statistics of each signal state block, then the set of SDPs described in Eq. (\ref{blockKeyRate}) help us reliably lower bound the key rate of the protocol.

\subsection{Source maps} \label{sectionSourceMaps}

We will now describe a commonly used class of source-replacement schemes which we call source maps, with ideas similar to squashing maps \cite{gittsovich2014squashing}. In general, source maps simplify security proofs at the cost of loosening our key rate bounds and giving Eve more power than she has in reality.
\begin{definition}[Source Map]
    Let $\{\rho_i\}\in\textrm{D}(\mathcal{H})$ and $\{\tau_i\}\in\textrm{D}(\mathcal{K})$ be the set of states Alice prepares for two QKD protocols where the rest of the protocol is the same. A channel $\Psi$ from $\textrm{D}(\mathcal{K})$ to $\textrm{D}(\mathcal{H})$ is a \textbf{source map} if $\rho_i = \Psi(\tau_i)$ for all $i$.
    We call the protocol where Alice produces the states $\{\rho_i\}$ ($\{\tau_i\}$) a real (virtual) protocol with real (virtual) states.
\end{definition}

\begin{figure}[ht]
    \begin{subfigure}{\linewidth}
        \centering
        \includegraphics[scale = 0.5]{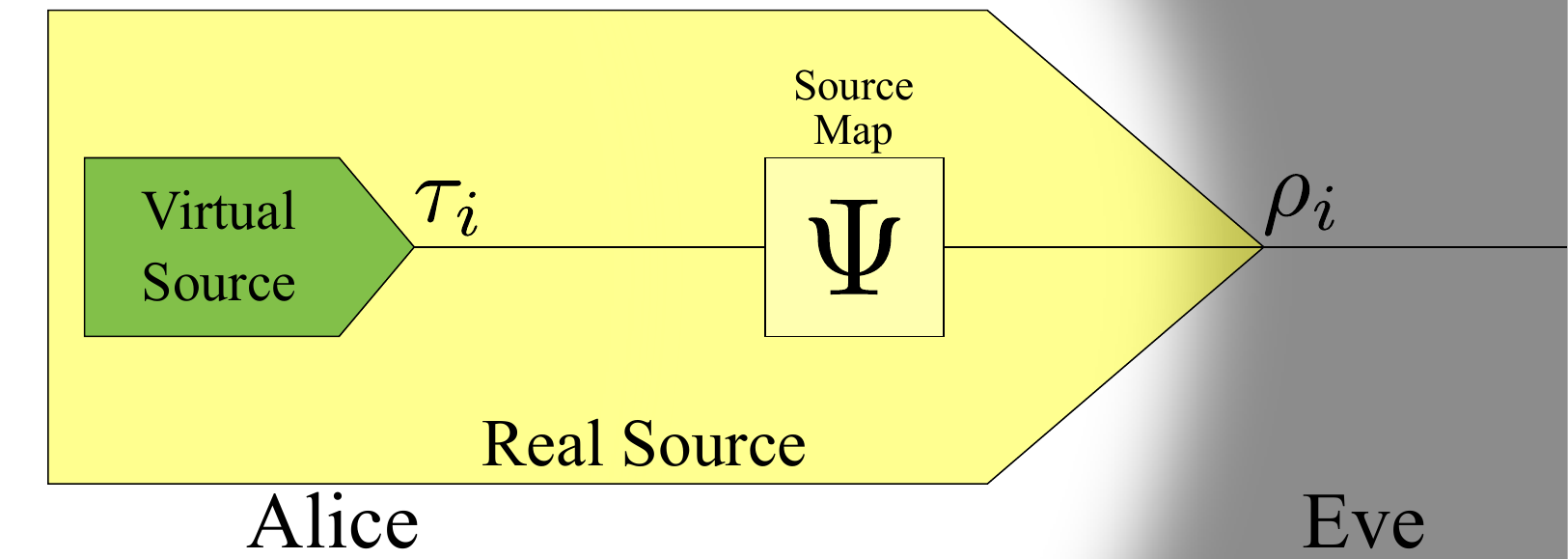}
        \caption{We can model the real source as a virtual source followed by a source map since they both have the exact same output.}  \label{figSourceMapBefore}
    \end{subfigure}
    \hfill
    \begin{subfigure}{\linewidth}
        \centering
        \includegraphics[scale= 0.5]{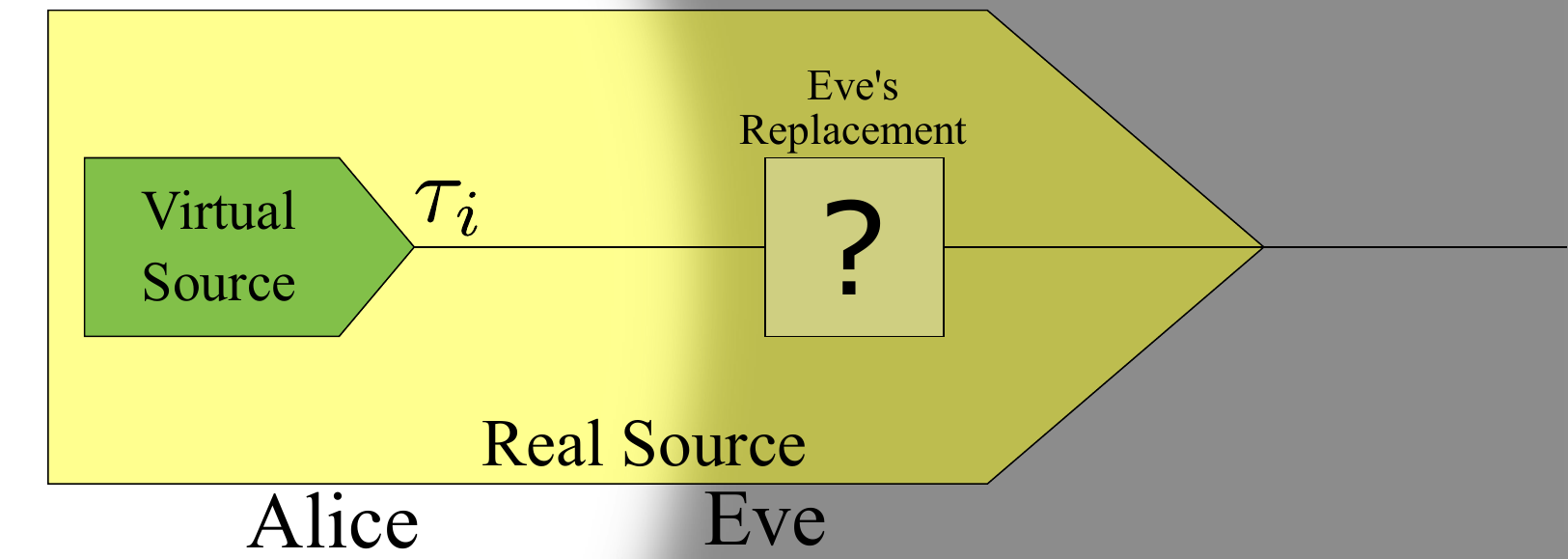}  
        \caption{Once we give Eve control of the source map, she can perform any physical operation on the output of the virtual source, including the source map $\Psi$ if reproducing the real state is optimal for her.}  \label{figSourceMapAfter}
    \end{subfigure}
    \caption{The real source can always be replaced by the real source in security proofs if they are related via a source map since the virtual source gives Eve more power.}  \label{figSourceMaps}
\end{figure}

Let $R^\infty_\rho$ and $R^\infty_\tau$ be the asymptotic key rates for identical observations $\gamma_{j\vert i}$ of the real and virtual QKD protocols respectively. The key rates are related as $R^\infty_\tau \leq R^\infty_\rho$. Intuitively, this can be seen from Fig. \ref{figSourceMaps} where giving Eve the source map gives her more power.
A more formal proof of this fact is given in Appendix \ref{appendixSourceMapKeyRateProof}.

Note that in the above definition, the different signal states $\{\rho_i\}$ and $\{\tau_i\}$ that Alice prepares could represent the joint state sent for multiple key generation rounds of the protocol. Thus, this does not assume either iid signal states or iid attacks by Eve, and is completely general.

As an example of a source map that we shall use, we describe virtual states call block-tagged states \cite{gottesman2004security}. Consider a protocol with an iid source that produces real states $\rho^{\mu_S}_i = V_i\rho^{\mu_S} V_i^\dag$ where $\rho^{\mu_S}$ can be diagonalised as $\rho^{\mu_S} = \sum_{\tilde{n}} p_{\tilde{n}} \ket{\tilde{n}}\bra{\tilde{n}}$. We can then define the virtual `block-tagged' states as $\tau_i = \sum_{\tilde{n}} p_{\tilde{n}} V_i\left( \ket{\tilde{n}}\bra{\tilde{n}}\right)V_i^\dag \otimes \ket{\tilde{n}}\bra{\tilde{n}}$, and the source map $\Psi = \mathbb{I} \otimes \textrm{Tr}$ that reproduces the real states from the virtual states is the partial trace over the second system. We call this simplification block-tagging.

The block-diagonal structure of the block-tagged states simplifies the objective function by breaking it up into individual blocks \cite{li2020improving} as $f(\rho^N) = \sum_{\tilde{n}} p_{\tilde{n}} f(\rho_{\tilde{n}}^N)$ where
\begin{align*}
    \rho_{\tilde{n}}^N = \sum_{i,j}& \sqrt{p(i)p(j)}\ket{i}\bra{j}_{A} \otimes\\ &\Pi_N\Phi\left(V_i\left(\ket{\tilde{n}}\bra{\tilde{n}}\right)V_j^\dag \otimes \ket{\tilde{n}}\bra{\tilde{n}}\right)_{B}\Pi_N.
\end{align*}
Thus, we can use Eq. (\ref{blockKeyRate}) to bound the key rate even if the real states do not have the block-diagonal structure. This simplification comes at the cost of key rate in the case that the isometries do not retain the block-diagonal structure of the state i.e. $\bra{\tilde{n}}V_i^\dag V_j \ket{\tilde{m}} \neq 0$ for some $\tilde{n}\neq\tilde{m}$. We note that a key rate simplification similar to Eq. (\ref{blockKeyRate}) was first seen in the context of discrete-phase-randomised decoy-state QKD in \cite{cao2015discrete}, although they use different techniques to arrive at the result.

\section{Phase imperfections in QKD} \label{secPhaseImperfections}

We shall first discuss a simplified model of phase imperfections that we consider. We then describe a source map that connects a model iid state to the imperfect state for a large class of QKD protocols.
Finally, we also discuss some of the difficulties in characterising the relevant parameters to construct the model iid state from the actual laser state.

We model a sequence of laser pulses as a probabilistic mixture of coherent states, where different laser pulses are independent of each other. Since we only consider phase imperfections, we assume that the intensity of each laser pulse is the same. Under these assumptions, the general state for the sequence of laser pulses can be written as
\begin{align} \label{generalLaser}
    \nonumber \rho^\mu_{\rm{laser}} =& \int\! d\phi_1\!\ldots\! d\phi_n\, p_{\Phi_1}(\phi_1) \ldots p_{\Phi_n}(\phi_n)\\
    & \ket{\sqrt{\mu} \rm{e}^{\rm{i}\phi_1}}\bra{\sqrt{\mu} \rm{e}^{\rm{i}\phi_1}}\otimes\ldots \otimes \ket{\sqrt{\mu} \rm{e}^{\rm{i}\phi_n}}\bra{\sqrt{\mu} \rm{e}^{\rm{i}\phi_n}}.
\end{align}

We will show that we can replace this general source state by a simplified state that is iid and is of the form
\begin{align} \label{modelLaser}
    \nonumber \left({\rho_{\rm{model}}^\mu}\right)^{\otimes n} = \,&q \int\! d\phi \, \frac{1}{(2\pi)^n}\ket{\sqrt{\mu} \rm{e}^{\rm{i}\phi}}\bra{\sqrt{\mu} \rm{e}^{\rm{i}\phi}}^{\otimes n}\\
    &+ (1-q) \ket{\sqrt{\mu}}\bra{\sqrt{\mu}}^{\otimes n}
\end{align}
where $q \coloneqq  \underset{k}{\min}\, \underset{\phi_k}{\min}\, 2\pi p_{\Phi_k}(\phi_k)$ is a parameter that must be characterised which represents the degree to which the sequence of laser pulses are phase-randomised. Although characterising this parameter might pose some practical difficulties, it is still significantly easier than characterising each probability density function $p_{\Phi_k}$.

\subsection{Source map for non-iid laser} \label{sectionSourceMapCorrLaser}

\begin{figure}[h]
    \centering
    \includegraphics[width= \linewidth]{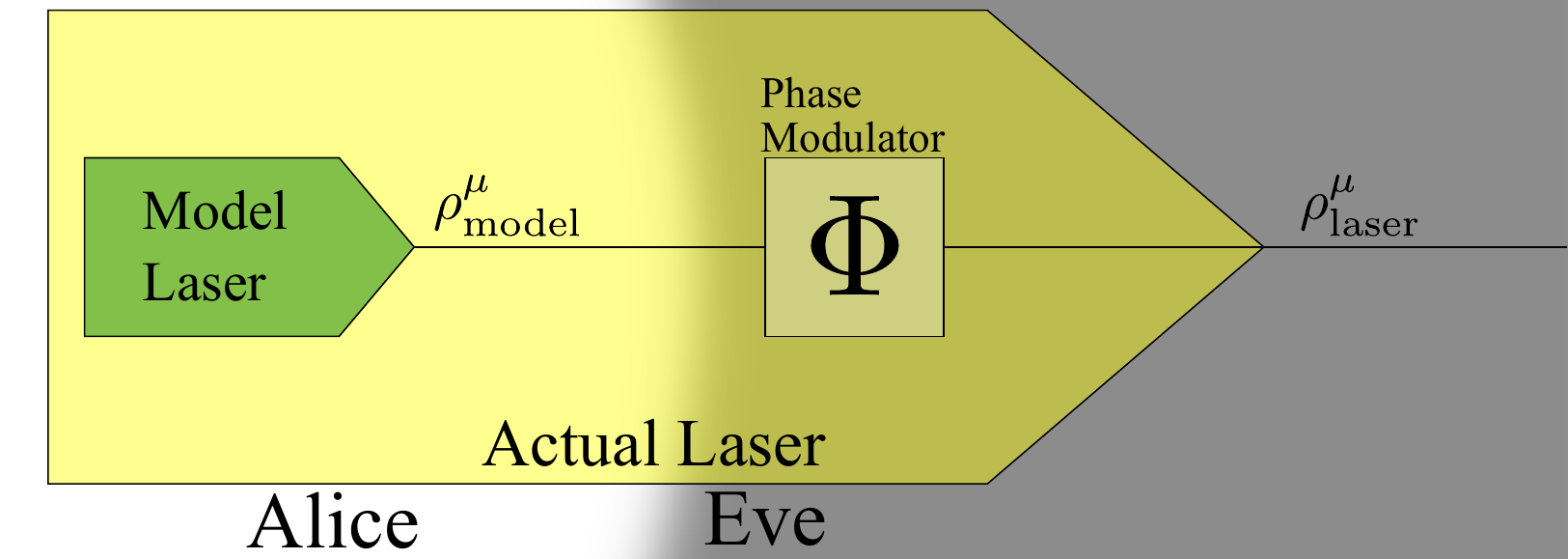}
    \caption{Replace the actual laser source with a model laser source and a phase modulator. We can then give the phase modulator to Eve. This would increase the power that Eve has and thus the key rate using the model laser source would lower bound the key rate using the actual laser source.}  \label{simplifiedModel}
\end{figure}

We now explicitly construct a physical map that connects the model laser state to the actual laser state with phase distribution $p_{\Phi_1\ldots \Phi_n}(\phi_1 \ldots \phi_n)=p_{\Phi_1}(\phi_1) \ldots p_{\Phi_n}(\phi_n)$ with associated parameter $q$ as shown in Fig. \ref{simplifiedModel}.
As a first step toward the source map construction, we consider the action of a phase modulator on the model laser state. The phase modulator modulates the phase of the $i^\textrm{th}$ pulse with probability $\frac{p_{\Phi_i}(\phi_i)-q/2\pi}{1-q}$ for all $i$.
The model laser source together with this phase modulator will imitate the actual laser source i.e. $\rho^\mu_\textrm{laser} = \Phi(\rho^\mu_\textrm{model})$ where $\Phi$ represents the action of the phase modulator as described above. We give a proof of this in Appendix \ref{appendixSourceMapForLaser}.

However the definition of the source map $\Psi$ requires that $\Xi_i^{\otimes n}(\rho^\mu_\textrm{laser}) = \Psi(\Xi_i(\rho^\mu_\textrm{model})^{\otimes n})$ for \textit{all} signal states $i$ where $\Xi_i$ denotes the preparation channel that acts on a single pulse to prepare the final signal state for a single round of the protocol. We can construct the source map $\Psi$ for a large class of QKD protocols analogously to how we constructed the map $\Phi$. Intuitively, this construction holds when the preparation channels $\Xi_i$ for the protocol "commute" with the action of the phase modulator $\Phi$.

For example, consider a QKD protocol that uses time-bin encoding $\Xi_i$ where a single laser pulse is split into a block of two pulses with possible phase coherences across pulses. We construct the source map $\Psi$ through the action of a phase modulator that modulates the phase of the laser pulses as follows:
the $i^\textrm{th}$ block of pulses (which all are the output of the action of $\Xi_i$ on the $i^\textrm{th}$ laser pulse from $\rho^\mu_\textrm{model}$) are all modulated with the same phase $\phi_i$ with probability $\frac{p_{\Phi_i}(\phi_i)-q/2\pi}{1-q}$ for all $i$.
Note that this source map can be naturally extended to blocks with more than two pulses.

The source map we constructed would commute with any intensity modulation of the laser pulse. So, this would also be a valid source map for decoy-state protocols.
Thus, the key rate of the virtual protocol with an iid characterised laser source and preparation channels $\Xi_i$ would lower bound the key rate of the real protocol with the partially characterised non-identically distributed laser source and preparation channels $\Xi_i$.

\subsection{Experimental characterisation of laser} \label{sectionCharLaser}

We have constructed a source map from an uncorrelated laser source with different probability density functions $p_{\Phi_i}$ for each pulse, all satisfying $p_{\Phi_i}(\phi_i) \geq \frac{q}{2\pi} \quad \forall i,\phi_i$. Thus, the experimental problem has been reduced from characterising the probability density function for each pulse, to characterising a single parameter $q$ that represents the degree of phase-randomisation. Although this problem is a significantly simpler problem to solve, standard visibility measurements do not directly measure this quantity.

\begin{figure}[ht]
    \centering
    \includegraphics[width = \linewidth]{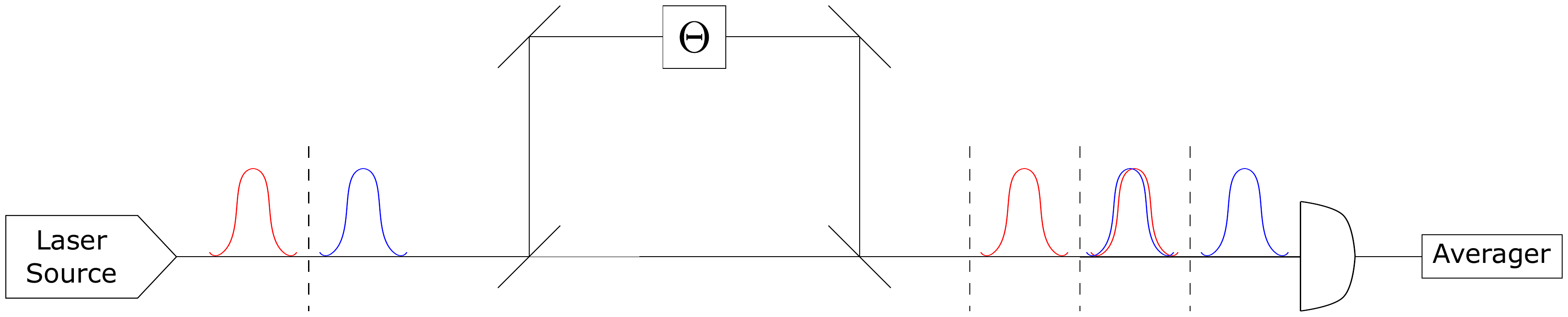}
    \caption{A schematic illustration of the experimental setup described in \cite{kobayashi2014evaluation}. The phase difference $\theta$ between the paths can be modulated with the phase shifter $\Theta$.} \label{expSetup}
\end{figure}

The visibility experiment as described in Section \textup{\uppercase\expandafter{\romannumeral2}}. A. of \cite{kobayashi2014evaluation}, is performed with a train consecutive laser pulses passed through an interferometer with a phase shifter in one arm. We have illustrated this with two pulses in Fig. \ref{expSetup}.
The intensity of the light arriving at the middle time slot of the detector is measured for different values of the phase shift $\theta$. We assume that the pulses have the same intensity.

The phase difference $\theta$ between the paths is varied to calculate the visibility $V$ given by
\begin{align}
    V =& \frac{I_{\max}-I_{\min}}{I_{\max}+I_{\min}}\\
    =& \frac{\langle\cos{(\theta+\phi)}\rangle_{\max}-\langle\cos{(\theta+\phi)}\rangle_{\min}}{2+\langle\cos{(\theta+\phi)}\rangle_{\max}+\langle\cos{(\theta+\phi)}\rangle_{\min}} \label{equationVisibility}
\end{align}
where the maximum and minimum is over all $\theta$, and $\phi$ is the difference in the phase of the adjacent pulses.

Note that the measurement as described in \cite{kobayashi2014evaluation} is used to measure phase correlations between adjacent pulses. However, due to limitations in our security proof techniques, we assume that the pulses are independent of each other. Moreover, the visibility measurement does not directly measure the degree of phase-randomisation $q$ directly as visibility measures other effects like the temporal distribution of the laser pulses.
Thus, using this experiment to obtain the extent of phase-randomisation $q$ requires us to make further model assumptions for the probability distribution.

As an illustrative example consider two different model assumptions for the phase distribution:
\begin{itemize}
    \item The phase distribution is the same as in the model laser state i.e. $p_{\Phi_i}(\phi_i) = \frac{q}{2\pi}+(1-q) \delta(\phi_i)$. We can then calculate
    \begin{align}
        q = 1-\sqrt{V}.
    \end{align}
    \item The phase distribution of each pulse is a wrapped normal distribution with standard deviation $\sigma$ centered about the origin i.e. $p_{\Phi_i}(\phi_i) = \frac{1}{\sqrt{2\pi\sigma^2}}\sum_{k=-\infty}^\infty \exp[\frac{-(\phi_i+2\pi k)^2}{2\sigma^2}].$ We can relate the visibility to the standard deviation as
    \begin{align}
        V = \exp[-\sigma^2].
    \end{align}
    As this completely characterises the wrapped normal distribution, we can use this to numerically find the extent of phase-randomisation $q$.
\end{itemize}
These values of $q$ computed under different model assumptions are, in general, different.
Thus, it would be interesting to develop other techniques that more directly measures this quantity and reduces the number of assumptions that we need to make.

To summarise, we have made two model assumptions on the laser state.
\begin{enumerate}
    \item The first model assumption is a limitation of our security proof techniques and can be stated as follows.
    The laser outputs a
    \begin{enumerate}
        \item probabilistic mixture of coherent states
        \item same intensity, and
        \item independent phase distribution. \label{modelAssumptionIndPhaseDist}
    \end{enumerate}
    Thus, the state can be written as
    \begin{align*}
        \rho^\mu_{\rm{laser}} =& \int\! d\phi_1\!\ldots\! d\phi_n\, p_{\Phi_1}(\phi_1) \ldots p_{\Phi_n}(\phi_n)\\
        & \ket{\sqrt{\mu} \rm{e}^{\rm{i}\phi_1}}\bra{\sqrt{\mu} \rm{e}^{\rm{i}\phi_1}}\otimes\ldots \otimes \ket{\sqrt{\mu} \rm{e}^{\rm{i}\phi_n}}\bra{\sqrt{\mu} \rm{e}^{\rm{i}\phi_n}}.
    \end{align*}
    Assumption \ref{modelAssumptionIndPhaseDist} has been lifted in \cite{curty2022phasedecoy}.
    \item The second assumption is due to limitations in the current experiments used to quantify the degree of phase-randomisation $q$. These further model assumptions on $p_{\Phi_i}(\phi_i)$ might be physically motivated. For eg.-
    \begin{enumerate}
        \item $p_{\Phi_i}(\phi_i) = \frac{q}{2\pi}+(1-q) \delta(\phi_i)$,
        \item $p_{\Phi_i}(\phi_i)$ is a wrapped normal distribution centered about the origin with standard deviation $\sigma$.
    \end{enumerate}
    Thus, it would be of practical interest to design new experiments to bound the minimum of the phase distribution without making such model assumptions.
\end{enumerate}

\section{Generalised Decoy-State Analysis} \label{sectionGenDecoy}

The standard decoy-state analysis relies on the assumption that the laser pulses are completely phase-randomised, hence block-diagonal. Additionally, it requires that the weight of each block is independent of the encoding used. However, the methods described in Section \ref{secPhaseImperfections} result in partially phase-randomised states of the form shown in Eq. (\ref{modelLaser}).
We first formulate the decoy-state problem abstractly by drawing an analogy to channel tomography in Section \ref{secGenDecoyGenFramework}.

For an iid fully phase-randomised source, we show in Section \ref{secStandDecoy} how the general formulation simplifies to the standard decoy-state analysis \cite{lo2005decoy,wang2005beating}. We stress the importance of the generalised decoy-state analysis for non-ideal sources as seen in Section \ref{secPhaseImperfections} since the standard decoy-state analysis cannot be used for laser states of the form described by Eq. (\ref{modelLaser}).

The general framework of our generalised decoy-state analysis typically takes the form of infinite-dimensional SDPs. In Section \ref{secFinProj} we introduce finite projections to construct a related finite-dimensional SDP that facilitates numerical evaluation. Finally, in Section \ref{secApptoDecoy} we describe a useful loosening of the SDP to reduce the dimensions while using it for typical QKD protocols.

\subsection{General framework} \label{secGenDecoyGenFramework}

First, to set up notation, let $\rho^k_\mu, \sigma_i \in \mathcal{D}(\mathcal{H})$ be density operators on $\mathcal{H}$ which we shall call the state space. Let $\Gamma_l, F_j \in \text{Pos}(\mathcal{K})$ be POVM elements on $\mathcal{K}$ which we shall call the measurement space. Let $\Phi: \mathcal{L}(\mathcal{H}) \xrightarrow[]{} \mathcal{L}(\mathcal{K})$ be a quantum channel.

We are given the statistics $\{\gamma_{l\vert k,\mu}\}$ of the input states $\{\rho^\mu_k\}$ to the unknown channel $\Phi$ where the output is measured by the POVM elements $\{\Gamma_l\}$ as $\Tr{\Gamma_l\Phi(\rho^\mu_k)} = \gamma_{l\vert k,\mu}$. We call these the actual states and POVM elements respectively. From this we seek to bound the statistics for a possibly different set of input states $\{\sigma_i\}$ and POVM elements $\{F_j\}$ measuring the output of the same channel $\Phi$ which can be written as $\Tr{F_j\Phi(\sigma_i)}$. We call these virtual states and POVM elements, and define a matrix $Y$ whose elements are the statistics $Y(i,j) = \Tr{F_j\Phi(\sigma_i)}$.

More formally, we are interested in the set $\mathbb{Y}$ of all matrices $Y$ with elements $Y(i,j) = \textrm{Tr}\left[\Phi(\sigma_{i}) F_j\right]$ with constraints on $\Phi$ given by
\begin{equation} \label{eq:maxGenDecoy}
    \begin{aligned}
        &\textrm{Tr}\left[\Phi(\rho_k^\mu) \Gamma_l\right] = \gamma_{l\vert k,\mu} \ \forall k,l,\mu \\
        & \Phi \textrm{ is CPTP.}
    \end{aligned}
\end{equation}
Note that here the different elements $Y(i,j)$ are not independent of each other for $Y \in \mathbb{Y}$. This makes it hard to find and use $\mathbb{Y}$. Thus, to make it easier to use, we define $Y^L(i,j) = \underset{Y\in\mathbb{Y}}{\inf } Y(i,j)$, and $Y^U(i,j) = \underset{Y\in\mathbb{Y}}{\sup } Y(i,j)$. These can now be independently written as the solution to the set of optimisation problems as follows:
\begin{equation} \label{eq:minGenDecoy}
    \begin{aligned}
        Y^{L}(i,j) = \underset{\Phi}{\textrm{min} } &\textrm{Tr}\left[\Phi(\sigma_{i}) F_j\right] \\
        \textrm{s.t. } &\textrm{Tr}\left[\Phi(\rho_k^\mu) \Gamma_l\right] = \gamma_{l\vert k,\mu} \ \forall k,l,\mu \\
        & \Phi \textrm{ is CPTP.}
    \end{aligned}
\end{equation}
\begin{equation}
    \begin{aligned}
        Y^{U}(i,j) = \underset{\Phi}{\textrm{max} } &\textrm{Tr}\left[\Phi(\sigma_{i}) F_j\right] \\
        \textrm{s.t. } &\textrm{Tr}\left[\Phi(\rho_k^\mu) \Gamma_l\right] = \gamma_{l\vert k,\mu} \ \forall k,l,\mu \\
        & \Phi \textrm{ is CPTP.}
    \end{aligned}
\end{equation}

This simplification is a relaxation of our initial problem to independent optimisations for each virtual state $i$ and outcome $j$. As a result of this relaxation, we might sometimes see counter-intuitive behaviour as illustrated by the following example. In the absence of this relaxation, we know that computing bounds for the sum of virtual POVM elements would be the same as computing and then summing the individual bounds. However, counter-intuitively solving these relaxed SDPs for sums of virtual POVM elements might lead to better bounds than solving then summing the optimal values of the individual SDPs. This is not a fundamental limitation as it does not affect the original optimisation. It is a direct consequence of the relaxation we have made to bound these statistics.

The optimisation problems described in Eq.(\ref{eq:maxGenDecoy}) and Eq. (\ref{eq:minGenDecoy}) can be reframed as SDPs by considering the Choi-Jamiolkowski isomorphism of the channel $J$
\begin{equation}
    \begin{aligned} \label{InfiniteDecoySDP}
        \underset{J}{\textrm{opt. }} &\Tr{(\sigma_{i}^T\otimes F_j) J} \\
        \textrm{s.t. } &\textrm{Tr}\left[({\rho_k^{\mu}}^T\otimes \Gamma_l) J\right] = \gamma_{l\vert k,\mu} &\forall k,l,\mu \\
        & J \geq 0 \\
        & \textrm{Tr}_\mathcal{K} \left[J\right] = \mathbb{I}_{\mathcal{H}}
    \end{aligned}
\end{equation}
where opt. indicates that we have to optimise the objective function to find both the maximum and the minimum as separate SDPs.
In order to simplify notation, let $S_\infty$ be the feasible set of the SDP i.e.
\begin{align} \label{Sinfty}
    \nonumber S_\infty \coloneqq \Big\{& J \in \mathcal{B}(\mathcal{H}\otimes \mathcal{K})\; \Big\vert \;
    \tr{J}{\mathcal{K}} = \mathbb{I}_\mathcal{H},\; J\geq 0, \\
    & \Tr{({\rho_k^{\mu}}^T\otimes \Gamma_l) J} = \gamma_{l\vert k,\mu} \quad\forall k,l,\mu \Big \}.
\end{align}

\subsection{Standard decoy} \label{secStandDecoy}

In the special case where the laser emits states $\rho_\textrm{PR}^\mu$ that are fully phase-randomised states with intensity $\mu$, we show how our general analysis given in Eq. (\ref{InfiniteDecoySDP}) reduces to the standard decoy-state analysis.

The actual states $\{\rho_k^\mu\}$ are obtained by the action of the preparation channels $\{\Xi_k\}$ on the fully phase-randomised laser state as $\rho_k^\mu = \Xi_k\left(\rho_\textrm{PR}^\mu\right)$. The virtual states that we can use in Eq. (\ref{InfiniteDecoySDP}) are the $n$-photon states for different encodings i.e. $\sigma_i^n = \Xi_i\left(\ket{n}\bra{n}\right)$. The crucial assumption here is that each of the actual states can be written as a classical mixture of the virtual states as
\begin{align}
    \rho_i^\mu &= \sum_n p_\mu(n) \Xi\left(\ket{n}\bra{n}\right) \nonumber\\
    &= \sum_n p_\mu(n) \sigma_i^n.
\end{align}
The actual POVM elements $\Gamma_j$ are obtained from the measurement setup. The virtual POVM elements whose outcomes we bound are the same as the actual POVM elements $F_j = \Gamma_j$.

With these definitions, we can rewrite the SDPs in Eq. (\ref{InfiniteDecoySDP}) as
\begin{equation}
    \begin{aligned} \label{StandardDecoySDP}
        \underset{J}{\textrm{opt. }} &\textrm{Tr}\left[\left(\Xi_i(\ket{n}\bra{n})^T\otimes \Gamma_j\right) J\right] \\
        \textrm{s.t. } &\textrm{Tr}\left[({\rho_k^{\mu}}^T\otimes \Gamma_l) J\right] = \gamma_{l\vert k,\mu} &\forall k,l,\mu \\
        & J \geq 0 \\
        & \textrm{Tr}_\mathcal{K} \left[J\right] = \mathbb{I}_{\mathcal{H}}.
    \end{aligned}
\end{equation}
The constraints in this case simplify as follows
\begin{align}
    \Tr{({\rho_k^{\mu}}^T\otimes \Gamma_l) J} &= \Tr{\Phi\left(\rho_k^\mu\right)\Gamma_l}\\
    &= \Tr{\Phi\left(\sum_n p_\mu(n)\Xi_k\left(\ket{n}\bra{n}\right)\right)\Gamma_l}\\
    &= \sum_n p_\mu(n)\Tr{\Phi\left(\Xi_k\left(\ket{n}\bra{n}\right)\right)\Gamma_l}\\
    &= \sum_n p_\mu(n)p(l\vert k,n)= \gamma_{l\vert k,\mu}
\end{align}
where $p(l\vert k,n)$ is the probability of a detection corresponding to the POVM $\Gamma_l$ given that Alice sent $n$ photons encoded with the preparation channel $\Xi_k$.

Noting that the objective function of the SDPs in Eq. (\ref{StandardDecoySDP}) can be written as $p(j\vert i,n)$, the SDPs simplify to the set of linear programs
\begin{equation}
    \begin{aligned}
        \underset{p}{\textrm{opt. }} & p(j|i,1)\\
        \textrm{s.t. } & \sum_n p_\mu(n)p(l\vert k,n)= \gamma_{l\vert k,\mu} & \forall k,l,\mu\\
        & 0 \leq p(l\vert k,n) \leq 1 & \forall k,l,n,\mu.
    \end{aligned}
\end{equation}

Having described the reduction of the generalised decoy-state analysis reduces to the standard decoy-state analysis, we remark on a subtle difference in their application to QKD.
The objective function in Eq. \ref{StandardDecoySDP} corresponds to $\Tr{(\sigma_i^T\otimes \Gamma_j) J}$. However, bounds on the finite projection $\Tr{(\sigma_{i}^T\otimes \Gamma_j^N) J}$ are needed in the key rate SDP shown in Eq. (\ref{blockKeyRate}). We would then need to use the dimension reduction method described in \cite{upadhyaya2021dimension}. In contrast, using the generalised decoy-state analysis we can directly choose the virtual POVM elements $F_j = \Gamma_j^N$ to estimate bounds on the statistics of the projected POVM elements.

\subsection{Finite projections} \label{secFinProj}

The set of SDPs described in Eq. (\ref{InfiniteDecoySDP}) are typically infinite-dimensional as in the case for optical setups. The problem of numerically finding bounds on infinite-dimensional SDPs when optimising over quantum states has been considered in \cite{upadhyaya2021dimension}. We use similar ideas to extend this analysis to SDPs where we optimise over quantum channels instead to find bounds on the set of SDPs described in Eq. (\ref{InfiniteDecoySDP}).

The idea is to construct a carefully chosen set of finite-dimensional SDPs whose optimal values can be related to the optimal values of the infinite-dimensional SDPs. Recall from Eq. (\ref{Sinfty}) the definition of the feasible set $S_\infty$ of the infinite-dimensional SDPs.
In subsections \ref{secFinProjPartTrace} and \ref{secExpValConst}, we construct a feasible set $S_{MN}$ of the finite-dimensional SDPs such that $S_{MN} \supseteq (\Pi_M\otimes\Pi_N)\;S_\infty\;(\Pi_M\otimes\Pi_N)$ for finite-dimensional projectors $\Pi_M$ and $\Pi_N$.
This condition is used in subsection \ref{secFiniteProjObjFunc} to relate the optimal value of the infinite-dimensional SDPs to the optimal values of the finite-dimensional SDPs.

We shall now proceed by considering a sequence of relaxations corresponding to each of the three constraints that define $S_\infty$. We add each constraint one by one so that each lemma only has the constraints needed to prove the required inclusion, till we finally construct $S_{MN}$ in Lemma \ref{lemmaSMN} such that $(\Pi_M\otimes\Pi_N)\;S_\infty\;(\Pi_M\otimes\Pi_N) \in S_{MN}$.

We begin with the positivity constraint, $J\geq 0$.
Note that projecting does not affect the positivity of an operator, as can be shown from the definition of positivity. Thus
\begin{align} \label{PosConst}
    (\Pi_M\otimes\Pi_N)\;\textrm{Pos}(\mathcal{H\otimes\mathcal{K}})\;(\Pi_M\otimes\Pi_N) \subseteq \textrm{Pos}(\mathcal{H}_M\otimes\mathcal{K}_N)
\end{align}
where $\mathcal{H}_M$ and $\mathcal{K}_N$ are the subspaces of $\mathcal{H}$ and $\mathcal{K}$ onto which $\Pi_M$ and $\Pi_N$ project, respectively. This gives the first relaxation.

\subsubsection{Partial trace constraint} \label{secFinProjPartTrace}

We refer to the constraint $\tr{J}{\mathcal{K}} = \mathbb{I}_\mathcal{H}$, and the corresponding modification described in this subsection as the partial trace constraint.
\begin{lemma} \label{LemmaPartialTrace}
    Let
    \begin{align*}
        \textrm{T}_\infty \coloneqq \Big\{ J\in \mathcal{B}(\mathcal{H}\otimes \mathcal{K})\; \Big\vert \; \tr{J}{\mathcal{K}} = \mathbb{I}_\mathcal{H},\; J\geq 0\Big\},\textrm{ and}\\
        \textrm{T}_{MN} \coloneqq \Big\{ J^{MN}\in \mathcal{B}(\mathcal{H}_M\otimes\mathcal{K}_N)\; \Big\vert \; \tr{J^{MN}}{\mathcal{K}} \leq \Pi_M,\\J^{MN}\geq 0 \Big\}.
    \end{align*}
    Then $(\Pi_M\otimes\Pi_N)\;\textrm{T}_\infty\;(\Pi_M\otimes\Pi_N) \subseteq \textrm{T}_{MN}$.
\end{lemma}
\begin{proof}
    Eq. (\ref{PosConst}) shows that $(\Pi_M\otimes \Pi_N)\,J\,(\Pi_M\otimes \Pi_N)\geq 0$ for any $J\in \textrm{T}_\infty$. Thus, it is sufficient to show that $\tr{(\Pi_M\otimes \Pi_N)\,J\,(\Pi_M\otimes \Pi_N)}{\mathcal{K}}\leq \Pi_M$ for any $J\in \textrm{T}_\infty$.
    Using Eqs. (4.95) - (4.99) from \cite{upadhyaya2021tools} we get that $\tr{J}{K}-\tr{(\mathbb{I}_\mathcal{H}\otimes\Pi_N)\,J\,(\mathbb{I}_\mathcal{H}\otimes\Pi_N)}{K}\geq 0$.
    For any element of $(\Pi_M\otimes\Pi_N)\;\textrm{T}_\infty\;(\Pi_M\otimes\Pi_N)$, we find
    \begin{align} \label{partialTraceConsEq}
        \nonumber & \tr{(\Pi_M\otimes \Pi_N)\,J\,(\Pi_M\otimes \Pi_N)}{\mathcal{K}}\\
        \nonumber&= \Pi_M\tr{(\mathbb{I}_\mathcal{H}\otimes \Pi_N)\,J\,(\mathbb{I}_\mathcal{H}\otimes \Pi_N)}{\mathcal{K}}\Pi_M\\
        \nonumber&\leq \Pi_M\tr{J}{\mathcal{K}}\Pi_M\\
        &= \Pi_M
    \end{align}
    where we have used the fact that $\tr{J}{\mathcal{K}}-\tr{(\mathbb{I}_\mathcal{H}\otimes\Pi_N)\,J\,(\mathbb{I}_\mathcal{H}\otimes\Pi_N)}{\mathcal{K}} \geq 0$ to obtain the inequality, and that $\tr{J}{\mathcal{K}} = \mathbb{I}_{\mathcal{H}}$ to get the final equality. This completes the proof.
\end{proof}

\subsubsection{Expectation value constraints} \label{secExpValConst}

We refer to the constraints
\begin{align}
    \Tr{({\rho_k^{\mu}}^T\otimes \Gamma_l) J} = \gamma_{l\vert k,\mu} \quad \forall k,l,\mu
\end{align}
and the corresponding modifications described in this subsection as the expectation value constraints. For these constraints, we proceed in two steps. We would first construct in Lemma \ref{lemmaExpValConstStateProj} a set belonging to $\mathcal{B}(\mathcal{H}_M\otimes\mathcal{K})$. We then construct the final finite-dimensional set on $\mathcal{B}(\mathcal{H}_M\otimes\mathcal{K}_N)$ in Lemma \ref{lemmaSMN}. Recall that we have considered $\mathcal{H}_M$ and $\mathcal{K}_N$ to be finite-dimensional spaces embedded in $\mathcal{H}$ and $\mathcal{K}$.

First, we set up some notation. Given a projection $\Pi_M$, we can define the off-diagonal blocks $H_k^\mu \coloneqq \rho_k^\mu - \rho_k^{\mu M} - \rho_k^{\mu \overline{M}}$ where $\rho_k^{\mu M} \coloneqq \Pi_M \rho_k^\mu \Pi_M$, and $\rho_k^{\mu \overline{M}} \coloneqq \overline{\Pi}_M \rho_k^\mu \overline{\Pi}_M$. We also define the weight $w_{kM}^\mu \coloneqq \Tr{\rho_k^{\mu\overline{M}}}$ of the $k^\textrm{th}$ input state outside the $\Pi_M$ projected subspace. This can be further used to define $\epsilon_{kM}^\mu \coloneqq \lambda_{kM}^\mu \sqrt{w_{kM}^\mu}$ which measures how ``big" the off-diagonal block is where $\lambda_{kM}^\mu \coloneqq \norm{\sqrt{\rho_k^{\mu M}}^g\Pi_M\rho_k^\mu\overline{\Pi}_M\sqrt{\rho_k^{\mu \overline{M}}}^g}_\infty$ and $A^g$ is the generalised inverse of $A$.
These definitions are used in the following lemma whose proof can be found in Appendix \ref{appendixLemmaExpValConstProof}.
\begin{restatable}{lemma}{lemmaExpValConstProj} \label{lemmaExpValConstStateProj}
    Define
    \begin{align} \label{expValConsStateProjEq}
        \nonumber \textrm{E}_{M} & \coloneqq \Big\{ J^{M} \in \mathcal{B}(\mathcal{H}_M\otimes \mathcal{K})\; \Big\vert \;
        \tr{J^{M}}{\mathcal{K}} \leq \Pi_M,\;\\
        & J^{M}\geq 0, \;
        \nonumber \Tr{({\rho_k^{\mu}}^T\otimes \Gamma_l) J^{M}} \leq \gamma_{l\vert k,\mu}+\epsilon^\mu_{kM},\\
        & \Tr{({\rho_k^{\mu}}^T\otimes \Gamma_l) J^{M}} \geq  \gamma_{l\vert k,\mu}-w_{kM}^\mu-\epsilon^\mu_{kM} \; \forall k,l,\mu\Big\}.
    \end{align}
    Then, $(\Pi_M\otimes\eyeMeas)\;\textrm{S}_\infty\;(\Pi_M\otimes\eyeMeas) \subseteq \textrm{E}_{M}$ where $\textrm{S}_\infty$ is as defined in Eq. (\ref{Sinfty}).
\end{restatable}

The last step to construct the set $\textrm{S}_{MN}$ requires that we estimate an additional quantity. We need to find bounds on the weight $W_{kN}^\mu$ of the transmitted state $\Phi(\rho_k^\mu)$ outside the $\Pi_N$ projected subspace defined as $\Tr{\Phi(\rho_k^\mu)\Pi_N} \geq 1-W_{kN}^\mu$. Equivalently, this can be written as
\begin{align} \label{eqWtOutsideJSinfty}
    \Tr{({\rho_k^{\mu}}^T\otimes \Pi_N) J} \geq 1-W_{kN}^\mu
\end{align}
where $J\in \textrm{S}_\infty$.
The method to find this bound is protocol dependent, and can be derived from the expectation value constraints on $J\in\textrm{S}_\infty$. As an example, we have described one such method to find the bound for the three-state protocol in Appendix \ref{appendixBoundW}.
 
This leads us to the explicit construction of $\textrm{S}_{MN}$ when the POVM elements $\Gamma_l$ commute with the projection $\Pi_N$.
\begin{restatable}{lemma}{lemmaSMN} \label{lemmaSMN}
    Let $[\Pi_N, \Gamma_l] = 0$ $\forall l$, and define
    \begin{align} \label{defSMN}
        \nonumber &\textrm{S}_{MN} \coloneqq \Big\{ J^{MN} \in \mathcal{B}(\mathcal{H}_M\otimes \mathcal{K_N})\; \Big\vert \;\\
        &\nonumber \tr{J^{MN}}{\mathcal{K}} \leq \Pi_M,\quad J^{MN}\geq 0, \;\\
        &\nonumber \Tr{({\rho_k^{\mu}}^T\otimes \Gamma_l) J^{MN}} \geq  \gamma_{l\vert k,\mu}-W_{kN}^\mu-w_{kM}^\mu-2\epsilon^\mu_{kM},\\
        & \Tr{({\rho_k^{\mu}}^T\otimes \Gamma_l) J^{MN}} \leq \gamma_{l\vert k,\mu}+\epsilon^\mu_{kM} \; \forall k,l,\mu\Big\}.
    \end{align}
    Then, $(\eyeState\otimes\Pi_N)\;\textrm{E}_M\;(\eyeState\otimes\Pi_N) \subseteq \textrm{S}_{MN}$ where $\textrm{E}_M$ is as defined in Eq. (\ref{expValConsStateProjEq}).
\end{restatable}

The proof of the above lemma can be found in Appendix \ref{appendixLemmaSMN}.
The following corollary is a direct consequence of Lemma \ref{lemmaExpValConstStateProj} and Lemma \ref{lemmaSMN}.
\begin{corollary} \label{corProjSinfinSMN}
    $(\Pi_M\otimes\Pi_N)\;\textrm{S}_\infty\;(\Pi_M\otimes\Pi_N) \subseteq \textrm{S}_{MN}$.
\end{corollary}
\begin{proof}
    \begin{align*}
        &(\Pi_M\otimes\Pi_N)\;\textrm{S}_\infty\;(\Pi_M\otimes\Pi_N) \\
        = &(\mathbb{I}_{\mathcal{H}}\otimes \Pi_N)\left((\Pi_M\otimes\mathbb{I}_{\mathcal{K}})\;\textrm{S}_\infty\;(\Pi_M\otimes\mathbb{I}_{\mathcal{K}})\right)(\mathbb{I}_{\mathcal{H}}\otimes \Pi_N)\\
        \subseteq & (\mathbb{I}_{\mathcal{H}}\otimes \Pi_N)\textrm{E}_M(\mathbb{I}_{\mathcal{H}}\otimes \Pi_N)\\
        \subseteq & \textrm{S}_{MN}
    \end{align*}
    where the first inclusion follows from Lemma 2, and the second follows from Lemma 3.
\end{proof}

\subsubsection{Objective function} \label{secFiniteProjObjFunc}

Having constructed $\textrm{S}_{MN}$, we now relate the objective function $\Tr{(\sigma_{i}^T\otimes F_j) J^{MN}}$ to $Y(i,j) = \Tr{(\sigma_{i}^T\otimes F_j) J}$ where $J^{MN} \in \textrm{S}_{MN}$ and $J\in S_\infty$. This subsection thus completes the construction of the finite-dimensional SDP and relates it to the infinite-dimensional SDP of interest as outlined at the start of Section \ref{secFinProj}.

We first define $w_{iM} \coloneqq \Tr{\sigma_i^{\overline{M}}}$, and $\lambda_{iM} \coloneqq \norm{\sqrt{\sigma_i^{M}}^g\Pi_M\sigma_i\overline{\Pi}_M\sqrt{\sigma_i^{\overline{M}}}^g}_\infty$ similar to the definitions at the start of Section \ref{secExpValConst}. Recall also that we defined $Y^U(i,j) = \underset{J\in \textrm{S}_\infty}{\max} \Tr{(\sigma_{i}^T\otimes F_j) J}$ and $Y^L(i,j) = \underset{J\in \textrm{S}_\infty}{\min} \Tr{(\sigma_{i}^T\otimes F_j) J}$.

First, we consider virtual POVM elements $F \in \mathcal{B}(\mathcal{K}_N)$ that live in the finite-dimensional subspace described by $\Pi_N$. This is indeed the case of interest for the key rate SDP described in Eq. (\ref{blockKeyRate}).
\begin{theorem} \label{thmFiniteSDPFiniteObjPOVM}
    Let $F_j \in \mathcal{B}(\mathcal{K}_N)$ be POVM elements such that $F_j = \Pi_N F_j \Pi_N$. Then
    \begin{align}
        Y^L(i,j) &\geq \underset{J^{MN}\in \textrm{S}_{MN}}{\min} \Tr{(\sigma_{i}^T\otimes F_j) J^{MN}} -\epsilon_{iM},\\
        Y^U(i,j) &\leq \underset{J^{MN}\in \textrm{S}_{MN}}{\max} \Tr{(\sigma_{i}^T\otimes F_j) J^{MN}} + w_{iM} + \epsilon_{iM}
    \end{align}
    where $\epsilon_{iM}\coloneqq \lambda_{iM} \sqrt{w_{iM}}$.
\end{theorem}
\begin{proof}
    Let $J^U$ and $J^L$ be the optimal operators in $\textrm{S}_\infty$ such that
    \begin{align}
        Y^U(i,j) &= \Tr{(\sigma_{i}^T\otimes F_j) J^U}\\
        Y^L(i,j) &= \Tr{(\sigma_{i}^T\otimes F_j) J^L}.
    \end{align}
    Noting that $\Pi_N F_j \Pi_N =F_j$, we infer from Lemma \ref{lemmaExpValConstStateProj} that
    \begin{align}
         \nonumber Y^L(i,j)\geq &\Tr{\left({\sigma_i}^T \otimes F_j\right) (\Pi_M\otimes\Pi_N)\;J^L\;(\Pi_M\otimes\Pi_N)}\\
         &-\epsilon_{iM}\\
         \nonumber Y^U(i,j) \leq &\Tr{\left({\sigma_i}^T\otimes F_j\right) (\Pi_M\otimes\Pi_N)\;J^U\;(\Pi_M\otimes\Pi_N)}\\
         &+w_{iM}+\epsilon_{iM}.
    \end{align}
    
    Corollary \ref{corProjSinfinSMN} implies that $(\Pi_M\otimes\Pi_N)\;J^L\;(\Pi_M\otimes\Pi_N) \in \textrm{S}_{MN}$, and $(\Pi_M\otimes\Pi_N)\;J^U\;(\Pi_M\otimes\Pi_N) \in \textrm{S}_{MN}$. Thus we get
    \begin{align}
        \nonumber &\underset{J^{MN}\in \textrm{S}_{MN}}{\min} \Tr{(\sigma_{i}^T\otimes F_j) J^{MN}}\\
        \leq &\Tr{\left({\sigma_i}^T \otimes F_j\right) (\Pi_M\otimes\Pi_N)\;J^L\;(\Pi_M\otimes\Pi_N)},\\
        \nonumber &\underset{J^{MN}\in \textrm{S}_{MN}}{\max} \Tr{(\sigma_{i}^T\otimes F_j) J^{MN}}\\
        \geq &\Tr{\left({\sigma_i}^T\otimes F_j\right) (\Pi_M\otimes\Pi_N)\;J^U\;(\Pi_M\otimes\Pi_N)}.
    \end{align}
    Chaining these inequalities completes the proof.
\end{proof}

Next, we consider the more general case where the POVM elements do not live in a finite-dimensional subspace.
We first use theorem \ref{thmFiniteSDPFiniteObjPOVM} to find the bound $\Tr{({\sigma_i}^T\otimes \Pi_N) J} \geq 1-W_{iN}$ by choosing $F_j = \Pi_N$ and numerically solving the finite-dimensional SDP $\underset{J^{MN}\in \textrm{S}_{MN}}{\min} \Tr{(\sigma_{i}^T\otimes \Pi_N) J^{MN}}$. This can be used to state the following, more general theorem.
\begin{theorem} \label{thmFiniteSDP}
    Let $F_j \in \mathcal{B}(\mathcal{K})$ be a POVM element such that $F_j = F_j^N + F_j^{\overline{N}}$. Then
    \begin{align}
        Y^L(i,j) &\geq \underset{J^{MN}\in \textrm{S}_{MN}}{\min} \Tr{(\sigma_{i}^T\otimes F_j) J^{MN}} -\epsilon_{iM},\\
        \notag Y^U(i,j) &\leq \underset{J^{MN}\in \textrm{S}_{MN}}{\max} \Tr{(\sigma_{i}^T\otimes F_j) J^{MN}} + W_{iN}\\
        & \phantom{\leq \underset{J^{MN}\in \textrm{S}_{MN}}{\max} \Tr{(\sigma_{i}^T\otimes F_j}} + w_{iM} + 2\epsilon_{iM}.
    \end{align}
\end{theorem}
\begin{proof}
    Similar to the proof of Theorem \ref{thmFiniteSDPFiniteObjPOVM}, pick $J^U$ and $J^L$ as the optimal operators in $\textrm{S}_\infty$ such that
    \begin{align}
        Y^U(i,j) &= \Tr{(\sigma_{i}^T\otimes F_j) J^U}\\
        Y^L(i,j) &= \Tr{(\sigma_{i}^T\otimes F_j) J^L}.
    \end{align}
    From Lemma \ref{lemmaSMN}, we can show that
    \begin{align}
         \nonumber Y^L(i,j)\geq &\Tr{\left({\sigma_i}^T \otimes F_j\right) (\Pi_M\otimes\Pi_N)\;J^L\;(\Pi_M\otimes\Pi_N)}\\
         &-\epsilon_{iM}\\
         \nonumber Y^U(i,j) \leq &\Tr{\left({\sigma_i}^T\otimes F_j\right) (\Pi_M\otimes\Pi_N)\;J^U\;(\Pi_M\otimes\Pi_N)}\\
         &+W_{iN}+w_{iM}+2\epsilon_{iM}.
    \end{align}
    
    The rest of the proof uses Corollary \ref{corProjSinfinSMN} and is identical to the proof of Theorem \ref{thmFiniteSDPFiniteObjPOVM}.
\end{proof}

Theorem \ref{thmFiniteSDPFiniteObjPOVM} and Theorem \ref{thmFiniteSDP} let us bound $Y(i,j)$ in terms of the solution to a finite-dimensional SDP. This can be done numerically.
We note here that this generalised decoy-state analysis is fairly general and can also be applied outside decoy-state QKD for eg.- to bound the statistics of cat states in \cite{curty2019simple} by sending fully phase-randomised states. 

\subsection{Application to decoy-state QKD} \label{secApptoDecoy}

We shall now detail how we can apply these methods to a general decoy-state QKD protocol. We also detail a protocol dependent relaxation that reduces dimensions for more efficient computation. To this end, consider a QKD protocol with signal states $\rho_i^{\mu_S}$ that are compatible with isometric preparation channels $\Xi_i$ as $\rho_i^{\mu_S} = \Xi_i\left( \rho^{\mu_S}\right)$. Assume that the base state $\rho^{\mu_S}$ can be diagonalised as
\begin{align} \label{eqBaseSignalStateDiagonalisation}
    \rho^{\mu_S} = \sum_{\tilde{n}} p_{\tilde{n}}\ket{\tilde{n}}\bra{\tilde{n}}.
\end{align}

We can block-tag these signal states with the eigenvectors $\ket{\tilde{n}}$ as described in Section \ref{sectionSourceMaps}. Our key rate optimisation then reduces to the SDP given in Eq. (\ref{blockKeyRate}).
As shown in Eq.(\ref{blockKeyRate}), we need to compute upper ($Y^U_{\tilde{n}}(i,j)$) and lower ($Y^L_{\tilde{n}}(i,j)$) bounds on $\Tr{\Gamma_j^N \Phi \left(\ket{\tilde{n}}\bra{\tilde{n}}\right)}$. This can be done directly by using the generalised decoy-state analysis described above. We choose the virtual states $\sigma_{i,\tilde{n}} = \Xi_i\left(\ket{\tilde{n}}\bra{\tilde{n}}\right)$, actual states $\{\rho_i^\mu\}$, actual POVM elements $\{\Gamma_j\}$, and virtual POVM elements $\{\Gamma_j^N\}$ for the analysis.
The set of finite-dimensional SDPs resulting from the generalised decoy-state analysis can be written as
\begin{equation}
    \begin{aligned} \label{finiteDecoySDP}
        \underset{J^{MN}}{\textrm{opt.}}\ & \Tr{({\sigma_{i,\tilde{n}}^{M}}^T\otimes F^N_j) J^{MN}}\\
        \textrm{s.t. }
        & \Tr{({\rho_k^{\mu M}}^T\otimes \Gamma^N_l) J^{MN}} \leq \gamma_{l\vert k,\mu}+\epsilon^\mu_{kM}\\
        &\Tr{({\rho_k^{\mu M}}^T\otimes \Gamma^N_l) J^{MN}} \geq  \gamma_{l\vert k,\mu}-W_{kN}^\mu\\
        & \phantom{\Tr{({\rho_k^{\mu M}}^T\otimes \Gamma^N_l) J^{MN}} \geq}-w_{kM}^\mu-2\epsilon^\mu_{kM} \forall k,l,\mu\\
        & J^{MN}\geq 0\\
        & \tr{J^{MN}}{\mathcal{K}} \leq \Pi_M
    \end{aligned}
\end{equation}
where we have an independent SDP for each actual state and POVM element indexed by $i,{\tilde{n}}$ and $j$ respectively.

In some cases, it is more convenient to perform a relaxed version of this generalised decoy-state analysis that does not involve the preparation channels $\Xi_i$ as follows. Consider the set of infinite-dimensional SDPs described in Eq. (\ref{InfiniteDecoySDP})
\begin{equation}
    \begin{aligned} \label{infiniteDecoySDPQKD}
        \underset{J}{\textrm{opt.}}\  &\Tr{\left(\left(\Xi_i\left(\ket{{\tilde{n}}}\bra{{\tilde{n}}}\right)\right)^T\otimes F_j\right) J} \\
        \textrm{s.t. } &\Tr{\left(\left(\Xi_k\left(\rho^\mu\right)\right)^T\otimes \Gamma_l\right) J} = \gamma_{l\vert k,\mu} &\forall k,l,\mu \\
        & J \geq 0 \\
        & \tr{J}{\mathcal{K}} = \mathbb{I}_\mathcal{H}
    \end{aligned}
\end{equation}
where we have made the dependence of the actual and virtual states on the preparation channels $\Xi_i$ explicit.
Recall that the constraints are equivalent to $J$ being the Choi isomorphism of a channel $\Phi$. So Eq. (\ref{infiniteDecoySDPQKD}) is equivalent to
\begin{equation}
    \begin{aligned}
        \underset{\Phi}{\textrm{opt.}}\  &\Tr{F_j\; \Phi\left(\Xi_i\left( \ket{\tilde{n}}\bra{\tilde{n}}\right)\right)} \\
        \textrm{s.t. } &\Tr{\Gamma_l\;\Phi\left(\Xi_k\left( \rho^\mu\right)\right)} = \gamma_{j\vert k,\mu} &\forall k,l,\mu \\
        & \Phi \textrm{ is CPTP.}
    \end{aligned}
\end{equation}

Two relaxations can now simplify these optimisation problems:
\begin{enumerate}
    \item Ignoring all constraints where $k\neq i$,
    \item Taking $\Phi' = \Phi \circ \Xi_i$ to be the new optimisation variable. Note that since the composition of two channels is also a channel, $\Phi'$ is also a channel.
\end{enumerate}
This expands the set being optimised over as we no longer fix $\Xi_i$ as can be seen by writing the resulting set of optimisation problems
\begin{equation}
    \begin{aligned}
        \underset{\Phi'}{\textrm{opt.}}\  &\Tr{F_j\; \Phi'\left( \ket{\tilde{n}}\bra{\tilde{n}}\right)} \\
        \textrm{s.t. } &\Tr{\Gamma_l\;\Phi'\left( \rho^\mu\right)} = \gamma_{j\vert i,\mu} &\forall l,\mu \\
        & \Phi' \textrm{ is CPTP.}
    \end{aligned}
\end{equation}
Since these relaxations expand the feasible set, the max (min) will be upper (lower) bounds of the original SDPs given in Eq. (\ref{infiniteDecoySDPQKD}).

Rewriting this as an SDP using the Choi matrix formalism we get
\begin{equation}
    \begin{aligned} \label{infiniteDecoySDPQKDRelaxed}
        \underset{J}{\textrm{opt.}}\  &\Tr{\left(\ket{{\tilde{n}}}\bra{{\tilde{n}}}^T\otimes F_j\right) J} \\
        \textrm{s.t. } &\Tr{\left({\rho^\mu}^T\otimes \Gamma_l\right) J} = \gamma_{l\vert i,\mu} &\forall l,\mu \\
        & J \geq 0 \\
        & \tr{J}{\mathcal{K}} = \mathbb{I}_\mathcal{H}.
    \end{aligned}
\end{equation}
Finally, use the results stated in Section \ref{secFinProj} to replace Eq. (\ref{infiniteDecoySDPQKDRelaxed}) with the finite-dimensional SDP
\begin{equation}
    \begin{aligned} \label{dimReducDecoySDP}
        Y_{\tilde{n}}(i,j) = \underset{J^{MN}}{\textrm{opt.}}\  &\Tr{\left(\sigma_{\tilde{n}}^M\otimes F_j^N\right)\,J^{MN}} \\
        \textrm{s.t. } &\Tr{({\rho^{\mu M}}^T\otimes \Gamma^N_l) J^{MN}} \leq \gamma_{l\vert i,\mu}+\epsilon^\mu_M \\
        &\Tr{({\rho^{\mu M}}^T\otimes \Gamma^N_l) J^{MN}}\geq \gamma_{l\vert i,\mu}-W_{iN}^\mu\\
        &\phantom{\Tr{({\rho^{\mu M}}^T\otimes \Gamma^N_l) J^{MN}}\geq}-w_M^\mu-2\epsilon_M^\mu \;\forall l,\mu \\
        & J^{MN} \geq 0\\
        & \tr{J^{MN}}{\mathcal{K}} \leq \Pi_M
    \end{aligned}
\end{equation}
where $\sigma_{\tilde{n}} \coloneqq \ket{\tilde{n}}\bra{\tilde{n}}$.

For some preparation channels, the dimension of the SDPs in Eq. (\ref{dimReducDecoySDP}) are smaller than the dimensions of the SDPs in Eq. (\ref{finiteDecoySDP}) for the same $w_{kM}^\mu = w_M^\mu$. An example where this is the case is the three-state protocol described in Section \ref{sec3StateProtocol}. Thus, it is sometimes advantageous to relax the problem to the more computationally tractable SDPs described in Eq. (\ref{dimReducDecoySDP}).

\section{Approximate diagonalisation} \label{approxDiag}

The eigendecomposition shown in Eq. (\ref{eqBaseSignalStateDiagonalisation}) is crucial for block-tagging and generalised decoy-state analysis. The eigenvalues $p_{\tilde{n}}$ are used in the objective function of Eq. (\ref{blockKeyRate}). The eigenvectors $\ket{\tilde{n}}$ are used in Eq. (\ref{blockKeyRate}) when determining $\rho_A^{\tilde{n}}$ for the partial trace constraint, and in Eq. (\ref{finiteDecoySDP}) or Eq. (\ref{dimReducDecoySDP}) when determining $\sigma_{i,\tilde{n}}$ or $\sigma_{\tilde{n}}$ respectively.

Unfortunately, the  eigendecomposition might be hard to find exactly as these are infinite-dimensional operators that cannot be numerically diagonalised. However, the eigendecomposition of a finite projection can be numerically found. This motivates the following definitions.
Let $\rho$ represent the infinite-dimensional density operator whose eigendecomposition we would like to estimate. Define $\rho' = \rho^\Pi + \rho^{\overline{\Pi}}$ where $\rho^\Pi = \Pi\rho\Pi$ and $\rho^{\overline{\Pi}} = \overline{\Pi}\rho\overline{\Pi}$ for some finite projection $\Pi$.

Note that $\rho^\Pi$ can be numerically diagonalised, and this would constitute a subset of the eigenvalues and eigenvectors of $\rho'$. How closely the eigendecomposition of $\rho'$ will estimate the eigendecomposition of $\rho$ depends on the choice of projection $\Pi$.
A useful choice of $\Pi$ would be one where the off-diagonal blocks are ``almost" 0 so that intuitively the eigendecomposition $\rho^\Pi$ is ``nearly" that of $\rho$. This is formalised in the following theorem whose proof is given in Appendix \ref{appendixEigenBound}.

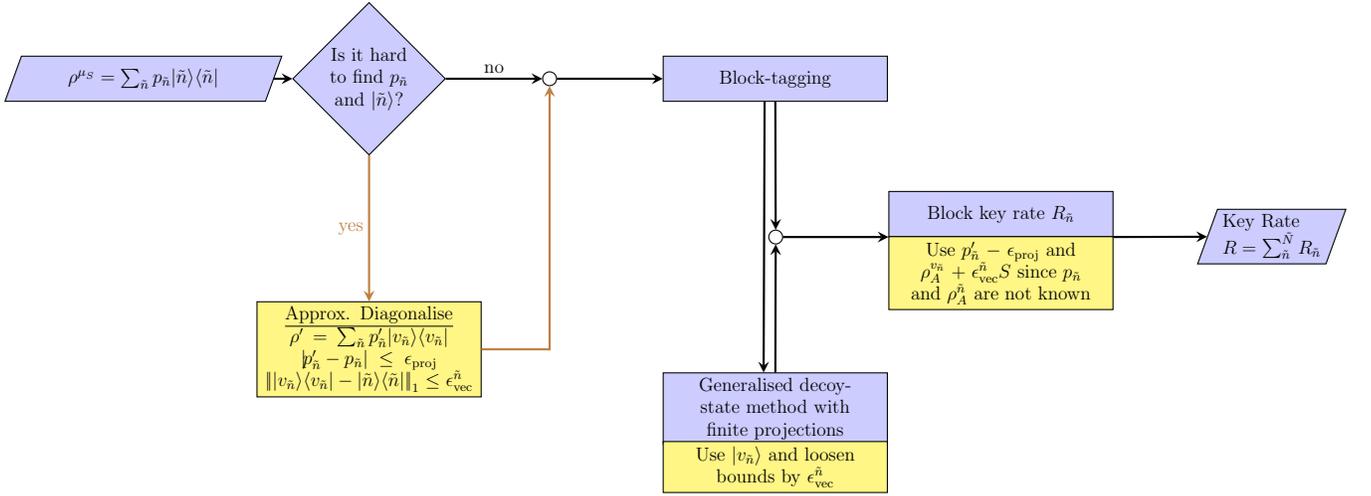
\begin{figure*}
    \centering
    \input{GeneralDecoyQKDFlowChart}
    \caption{Flowchart depicting the application of the generalised decoy-state method to QKD. The yellow parts denote the modifications to be made if we need to approximately diagonalise the density operator.} \label{genDecoyQKDFlowChart}
\end{figure*}

\begin{theorem} \label{thmApproxDiag}
    Let $\rho = \sum_{\tilde{n}} p_{\tilde{n}} \ket{\tilde{n}}\bra{\tilde{n}}$ where $p_0 \geq p_1 \geq\ldots$, and $\rho'$ have eigendecomposition
    \begin{align} \label{eqProjectedRhoEigendecomposition}
        \rho' = \sum_{\tilde{n}} p'_{\tilde{n}}\ket{v_{\tilde{n}}}\bra{v_{\tilde{n}}}
    \end{align}
    where $p'_0 \geq p'_1 \geq\ldots$. Define $\delta_{\tilde{n}} \coloneqq \textrm{min}\{p'_{\tilde{n}}-p'_{\tilde{n}-1}-\epsilon_\textrm{proj},\ p'_{\tilde{n}+1}-p'_{\tilde{n}}-\epsilon_\textrm{proj}\}$ where $\epsilon_{\textrm{proj}} \coloneqq \norm{\sqrt{\rho^\Pi}^g \Pi\rho\overline{\Pi} \sqrt{\rho^{\overline{\Pi}}}^g}_\infty \sqrt{\Tr{\rho^{\overline{\Pi}}}}$.
    Then
    \begin{enumerate}
        \item $\abs{\,p'_{\tilde{n}}-p_{\tilde{n}}} \leq \epsilon_\textrm{proj}$, and
        \item $F(\ket{v_{\tilde{n}}}\bra{v_{\tilde{n}}},\ket{\tilde{n}}\bra{\tilde{n}})^2 \geq 1-\frac{\epsilon_\textrm{proj}^2}{\delta_{\tilde{n}}^2}$.
    \end{enumerate}
\end{theorem}
Using Fuchs-van de Graaf inequality \cite{watrous2018theory} along with Theorem \ref{thmApproxDiag}, we get
\begin{align} \label{eqOneNormApproxDiag}
    \norm{\ket{v_{\tilde{n}}}\bra{v_{\tilde{n}}}-\ket{\tilde{n}}\bra{\tilde{n}}}_1 \leq 2\epsilon_\textrm{proj}/\delta_{\tilde{n}}.
\end{align}
For notational convenience, we define this quantity to be $\epsilon_\textrm{vec}^{\tilde{n}} \coloneqq 2\epsilon_\textrm{proj}/\delta_{\tilde{n}}$.
We can use Theorem \ref{thmApproxDiag} for QKD to approximately diagonalise $\rho^{\mu_S}$ as defined in Eq. (\ref{eqBaseSignalStateDiagonalisation}). This approximate diagonalisation would lead to minor modifications to the the generalised decoy-state bounds, as well as the key rate SDP as depicted in Fig. \ref{genDecoyQKDFlowChart}.

\subsection{Approximate generalised decoy-state analysis}

We first describe the use of Eq. (\ref{eqOneNormApproxDiag}) in obtaining bounds on the generalised decoy-state SDP given in Eq. (\ref{finiteDecoySDP}) where $\ket{\tilde{n}}$ appears in the objective function. By numerically diagonalising $\Pi\rho^{\mu_S}\Pi$, we can find $\sigma_{v_{\tilde{n}}} = \ket{v_{\tilde{n}}}\bra{v_{\tilde{n}}}$. This can be used instead of $\ket{\tilde{n}}$ to construct the virtual states for the objective function in Eq. (\ref{finiteDecoySDP}). Let the optimal values of the modified SDP be denoted by $Y_{v_{\tilde{n}}}(i,j)$. The optimal values of the original SDPs $Y_{\tilde{n}}(i,j)$ can be related to the optimal values of the modified SDPs $Y_{v_{\tilde{n}}}(i,j)$ by using the result in Eq. (\ref{eqOneNormApproxDiag}) with H\"older's inequality to get
\begin{align} \label{eqApproxDecoyObj}
    \abs{\,Y_{\tilde{n}}(i,j)-Y_{v_{\tilde{n}}}(i,j)} \leq \epsilon_\textrm{vec}^{\tilde{n}}.
\end{align}

Recall that the generalised decoy-state analysis makes use of finite projections $\Pi_M$ on the virtual state $\sigma_{v_{\tilde{n}}}$. This results in an additional $\epsilon_{iM}$ cost as described in Theorem \ref{thmFiniteSDPFiniteObjPOVM}. By choosing $\Pi\geq\Pi_M$, we can ensure that $\sigma_{v_{\tilde{n}}}=\Pi_M\sigma_{v_{\tilde{n}}}\Pi_M$ resulting in a reduced cost $\epsilon_{iM} = 0$. Thus, this suggests prudent choices for the different finite projections used in our analysis.

\subsection{Approximate key rate SDP}

The eigenvectors $\ket{\tilde{n}}$ appear in the key rate SDP in Eq. (\ref{blockKeyRate}) in the partial trace constraint $\tr{\rho_{AB}^{\tilde{n}N}}{B} \leq \rho^{\tilde{n}}_{A}$, and in the bounds $Y_{\tilde{n}}(i,j)$. The eigenvalues $p_{\tilde{n}}$ appear in Eq. (\ref{blockKeyRate}) as a prefactor to the objective function. We use the approximate eigenvectors $\ket{v_{\tilde{n}}}$ and eigenvalues $p_{\tilde{n}}$ to construct a similar SDP that bounds the key rate.
\begin{corollary}
    \begin{equation}
        \begin{aligned} \label{approxDiagKeyRateSDP}
            R_{\tilde{n}}^N \geq \underset{\rho^{\tilde{n} N}_{AB}}{\min}\;  & (p'_{\tilde{n}} - \epsilon_\textrm{proj}) f(\rho^{\tilde{n} N}_{AB})\\
            \textrm{s.t. } &  Y_{v_{\tilde{n}}}^L(i,j)-\epsilon_\textrm{vec}^{\tilde{n}}\leq \Tr{\Gamma_j^N\Phi(\rho^{\tilde{n}}_i)}\leq Y_{v_{\tilde{n}}}^U(i,j)+\epsilon_\textrm{vec}^{\tilde{n}}\\
            & \tr{\rho_{AB}^{\tilde{n}N}}{B} \leq \rho_{A}^{v_{\tilde{n}}}+\epsilon_\textrm{vec}^{\tilde{n}} S\\
            & \norm{S}_1 \leq 1\\
            & 1-W_{\tilde{n}}-\epsilon_\textrm{vec}^{\tilde{n}} \leq \Tr{\rho_{AB}^{\tilde{n}N}}\leq 1\\
            & S \geq 0\\
            & \rho_{AB}^{\tilde{n} N}\geq 0
        \end{aligned}
    \end{equation}
    where $R_{\tilde{n}}^N$ is defined in Eq. (\ref{blockKeyRate}).
\end{corollary}
\begin{proof}
    We first prove that any feasible $\rho^{\tilde{n} N}_{AB}$ for the SDP in Eq. (\ref{blockKeyRate}) is also feasible for the SDP in Eq. (\ref{approxDiagKeyRateSDP}).
    That $$Y_{v_{\tilde{n}}}^L(i,j)-\epsilon_\textrm{vec}^{\tilde{n}}\leq \Tr{\Gamma_j^N\Phi(\rho^{\tilde{n}}_i)}\leq Y_{v_{\tilde{n}}}^U(i,j)+\epsilon_\textrm{vec}^{\tilde{n}}$$ is implied by $$Y_{{\tilde{n}}}^L(i,j)\leq \Tr{\Gamma_j^N\Phi(\rho^{\tilde{n}}_i)}\leq Y_{{\tilde{n}}}^U(i,j)$$ is a direct consequence of Eq. (\ref{eqApproxDecoyObj}).
    
    Given that $\tr{\rho_{AB}^{\tilde{n}N}}{B} \leq \rho^{\tilde{n}}_{A}$, we aim to show that
    \begin{align*}
        \tr{\rho_{AB}^{\tilde{n}N}}{B} \leq \rho^{v_{\tilde{n}}}_{A} +\epsilon_\textrm{vec}^{\tilde{n}} S
    \end{align*}
    where $S$ is a positive semidefinite operator with $\norm{S}_1 \leq 1$.
    Recall from Eq. (\ref{eqPsiAAsA'}) that
    \begin{align}
        \ket{\psi^{\tilde{n}}}_{AA'} = \sum_{i} \sqrt{p(i)} \ket{i}_A \otimes V_i\ket{\tilde{n}}_{A'},\\ 
        \ket{\psi^{v_{\tilde{n}}}}_{AA'} = \sum_{i} \sqrt{p(i)} \ket{i}_A \otimes V_i\ket{v_{\tilde{n}}}_{A'} \label{eqApproxPsiAA'ntilde}
    \end{align}
    where $V_i$ is the isometry that define the isometric preparation channels $\Xi_i$. As a direct consequence of Theorem \ref{thmApproxDiag} we get
    \begin{align}
        F(\rho_{AA'}^{\tilde{n}},\rho_{AA'}^{v_{\tilde{n}}}) &= F(\ket{v_{\tilde{n}}}\bra{v_{\tilde{n}}},\ket{\tilde{n}}\bra{\tilde{n}})\\ 
        &\geq \sqrt{1-\frac{\epsilon_\textrm{proj}^2}{\delta_{\tilde{n}}^2}}
    \end{align}
    where $\rho_{AA'}^{\tilde{n}} = \ket{\psi^{\tilde{n}}}\bra{\psi^{\tilde{n}}}_{AA'}$ and $\rho_{AA'}^{v_{\tilde{n}}} = \ket{\psi^{v_{\tilde{n}}}}\bra{\psi^{v_{\tilde{n}}}}_{AA'}$.
    
    Thus, Fuchs-van de Graaf inequality can be used to obtain
    \begin{align}
        \norm{\rho_{AA'}^{\tilde{n}}-\rho_{AA'}^{v_{\tilde{n}}}}_1 \leq \epsilon_\textrm{vec}^{\tilde{n}}.
    \end{align}
    Since the partial trace channel can only decrease the one-norm, this gives
    \begin{align}
        \norm{\rho_{A}^{\tilde{n}}-\rho_{A}^{v_{\tilde{n}}}}_1 \leq \epsilon_\textrm{vec}^{\tilde{n}}.
    \end{align}
    Thus, the partial trace constraint $\tr{\rho_{AB}^{\tilde{n}N}}{B} \leq \rho^{\tilde{n}}_{A}$ implies
    \begin{align}
        \tr{\rho_{AB}^{\tilde{n}N}}{B} \leq \rho^{v_{\tilde{n}}}_{A} +\epsilon_\textrm{vec}^{\tilde{n}} S
    \end{align}
    where $S$ is a positive semidefinite operator with $\norm{S}_1 \leq 1$.
    
    Thus, the feasible set for the SDP in Eq. (\ref{approxDiagKeyRateSDP}) contains the feasible set for the SDP in Eq. (\ref{blockKeyRate}).
    Finally, Theorem \ref{thmApproxDiag} states that $$ p_{\tilde{n}}\geq p'_{\tilde{n}} - \epsilon_{\textrm{proj}}$$ completing the proof.
\end{proof}

\section{Three-state protocol} \label{sec3StateProtocol}

We shall now apply the methods developed so far to analyse the effects of imperfect phase-randomisation on the key rate of the time-bin encoded three-state protocol. This protocol can be implemented primarily by using passive components which are easy to manufacture. A recent implementation \cite{boaron2018secure} was able to share secret keys over 421 km under the assumption that the laser is fully phase-randomised. However, the 2.5 GHz laser used in the implementation did not perfectly randomise the phase \cite{grunenfelder2020performance} highlighting the importance of the methods developed in this paper.
 
\subsection{Protocol description} \label{secThreeStateProtocolDescription}

\subsubsection{State preparation}

Alice produces a laser pulse with some phase distribution as described in Eq. \ref{generalLaser}. She then passes it through an unbalanced Michelsons interferometer that transforms the coherent state from $\ket{\alpha} \rightarrow \ket{\alpha/2}\otimes \ket{\alpha/2}$. Alice randomly chooses a bit to encode from \{$0$, $1$, $+$\} with an a priori probability distribution and transforms the state accordingly:
\begin{itemize}
    \item[] $0$: Alice uses an intensity modulator to suppress the first pulse.
    \item[] $1$: Alice uses an intensity modulator to suppress the second pulse.
    \item[] $+$: Alice uses a variable attenuator to halve the intensity of each pulse so that the total mean photon number of both pulses in all 3 states are the same.
\end{itemize}
Additionally Alice uses the variable attenuator to send some decoy states with different intensities with the same encoding as the signal states.

\subsubsection{Measurement} \label{secMeas}

\begin{figure*}[t]
    \centering
    \includegraphics[width = \linewidth]{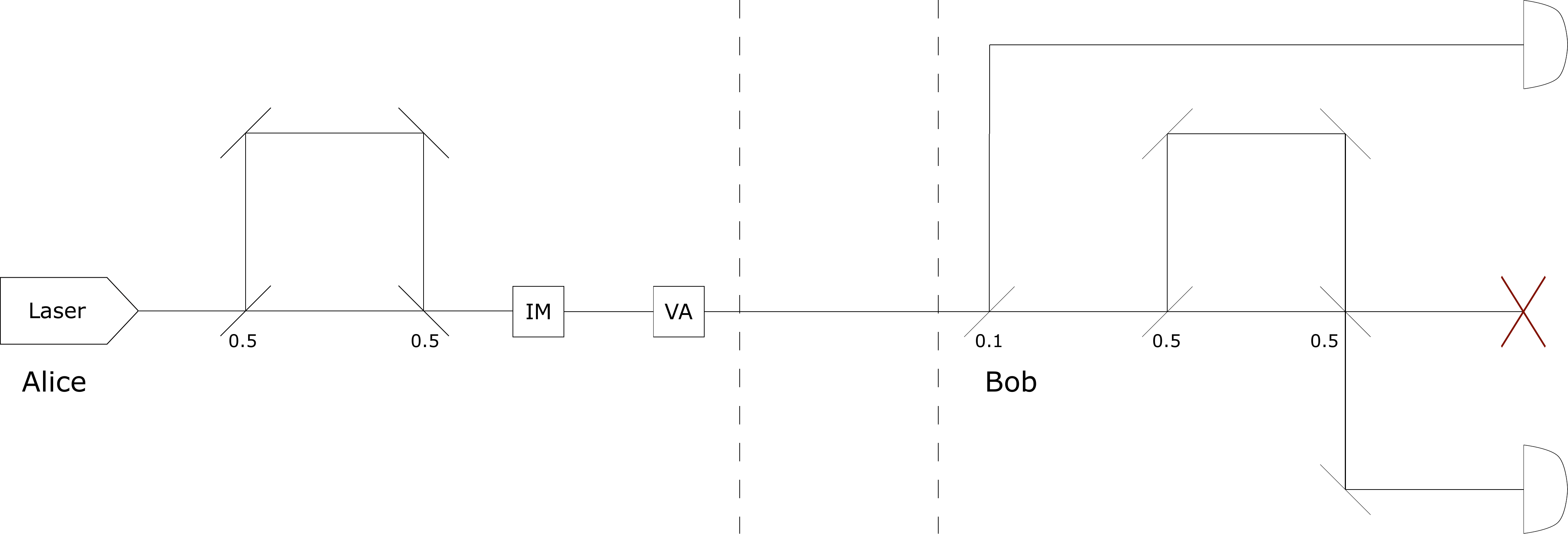}
    \caption{Schematics of the implementation of the three-state protocol as in \cite{boaron2018secure}. The numbers below the beam-splitters reflect their transmissivity. IM and VA refer to intensity modulator and variable attenuator respectively.} \label{3StateExpSetup}
\end{figure*}

Bob's basis choice is made passively via a beam-splitter. The Z basis detection is made by a threshold detector that measures the time of arrival. This measurement is used for key generation. The X basis detection is made via  a Mach-Zehnder interferometer that measures the coherences between pulses. Here, only the `-' detector is used for experimental simplicity. The setup is shown in Fig. \ref{3StateExpSetup}.

\subsubsection{Simulation parameters} \label{sectionParameters}

The laser visibility was measured \cite{grunenfelder2020performance} to be $V=0.0019$. In order to interpret this measurement result as the degree of phase-randomisation $q$, we need to make model assumptions on the general laser state as discussed in Section \ref{sectionCharLaser}. For the two physical model assumptions discussed in Section \ref{sectionCharLaser}, we get $q \approx 0.9564$ when we assume $p_{\Phi_i}(\phi_i) = \frac{q}{2\pi}+(1-q) \delta(\phi_i)$, and $q \approx 0.9128$ when $p_{\Phi_i}(\phi_i)$ is a wrapped normal distribution.

The channel is modelled as a loss-only channel with a low attenuation of $0.16$ dB/km based on the implementation in \cite{boaron2018secure}. We reduce the number of constraints to speed up computation time. In particular, we consider no click events, single click events, and group all multi-click events together as a single event. Bob's threshold detectors are assumed to be ideal without any dark counts or loss.

Ideally, we would want to optimise over all free parameters to maximise the key rate we can produce. However, this is computationally very taxing and so we pick some fixed arbitrary values for the free parameters. Thus, our results deliver provable secure key rates, but we do not claim optimality. Alice's states are all chosen with equal a priori probabilities. The decoy amplitudes used are 0 and 0.5 while the signal intensity is optimised for different distances. Bob's passive beam-splitter is a 0.9/0.1 beam-splitter with the 0.9 being towards the Z basis choice.

\subsection{Applying generalised decoy-state analysis}

The laser is characterised as described in Section \ref{sectionCharLaser} to obtain values for the degree of phase-randomisation $q$. Using this parameter with the source map described in Section \ref{sectionSourceMapCorrLaser} we reduce the general problem to finding the key rate given Alice's prepared states $\rho_i^\mu = \Xi_i\left( \rho^\mu_\textrm{model} \right)$. Note that since the three states have the same mean photon number, the preparation channels $\Xi_i$ can be represented by isometric channels by choosing the base state to also have the same mean photon number.
We can now follow the process depicted in Fig. \ref{genDecoyQKDFlowChart} to obtain the key rate for these states.

This first step is to approximately diagonalise $\rho^{\mu_S}_\textrm{model}$ where $\mu_S$ is the signal intensity. We take a finite projection in photon number space upto $d$ photons $\Pi_d$ to numerically diagonalise the operator. Let $\ket{v_{\tilde{n}}}$ and $\lambda_{\tilde{n}}$ be the resulting eigenvectors and eigenvalues. Using Theorem IX.5.9 from \cite{bhatia2013matrix} along with the fact that
\begin{align}
    \nonumber \rho_{\rm{model}}^{\mu_S}= &q \sum_{n=0}^\infty e^{-\mu_S} \frac{\mu_S^n}{n!} \ket{n}\bra{n}\\
    &+ (1-q) \ket{\sqrt{\mu_S}}\bra{\sqrt{\mu_S}} \label{eqRhoModel3State}
\end{align}
is positive semidefinite for all $q \in [0,1]$, we can conclude that 
\begin{align*}
    \norm{\sqrt{\Pi_d\rho_\textrm{model}^{\mu_S}\Pi_d}^g\Pi_d\rho_\textrm{model}^{\mu_S}\overline{\Pi}_d\sqrt{\overline{\Pi}_d\rho_\textrm{model}^{\mu_S} \overline{\Pi}_d}^g}_\infty \leq (1-q).
\end{align*}
$w_d^{\mu} = \Tr{\rho^{\mu}\overline{\Pi}_d}$ is given by $w_d^{\mu} = 1-\sum_{n=0}^d e^{-\mu} \frac{\mu^n}{n!}.$
The relevant bounds $\epsilon_\textrm{proj}^{\mu_S}$ and ${\epsilon_\textrm{vec}^{{\mu_S\tilde{n}}}}$ can then be found for the state $\rho^{\mu_S}_\textrm{model}$ to use the results shown in Section \ref{approxDiag}.
We can also block-tag the signal state given in Eq. (\ref{eqRhoModel3State}) as shown in Section \ref{sectionSourceMaps} to choose the relevant virtual states for the decoy-state analysis.

We can then use the generalised decoy-state SDP as described in Section \ref{secApptoDecoy}. In order to save computational time, we use the reduction given in Eq. (\ref{dimReducDecoySDP}). We choose to project onto the space with less than or equal to $N$ photons in both pulses $\Pi_N$ when considering the measurement space. This commutes with all the POVM elements $\Gamma_j$ since they are all threshold detectors. The projection on the state space $\Pi_M$ is chosen to be the same as the projection used for approximate diagonalisation $\Pi_d$. Thus, $\Pi_M\leq\Pi_d$ and $\sigma_{\tilde{n}} = \ket{v_{\tilde{n}}}\bra{v_{\tilde{n}}}$ already lives entirely within the space spanned by $\Pi_M$. So the correction term described in Theorem \ref{thmFiniteSDPFiniteObjPOVM} goes to 0. Additionally, $w_M^\mu = w_d^\mu$.

To apply Eq. (\ref{dimReducDecoySDP}) the quantities $\epsilon_M^\mu$ and $W_{iN}^\mu$ still need to be computed. $\epsilon_M^\mu$ can be bound as described in Lemma \ref{lemmaExpValConstStateProj} as $\epsilon_M^\mu \leq (1-q) w_M^\mu$. It is possible to bound $W_{iN}^\mu$ from our observations as shown in Appendix \ref{appendixBoundW}.
This fully defines the finite dimensional SDPs shown in Eq. (\ref{dimReducDecoySDP}) that can be numerically solved. The solutions of these SDPs together with the other parameters can be used to define the key rate SDP shown in Eq. (\ref{approxDiagKeyRateSDP}). This gives us a lower bound on the key rate.

\subsection{Results}

\begin{figure}[ht]
    \centering
    \includegraphics[width = \linewidth]{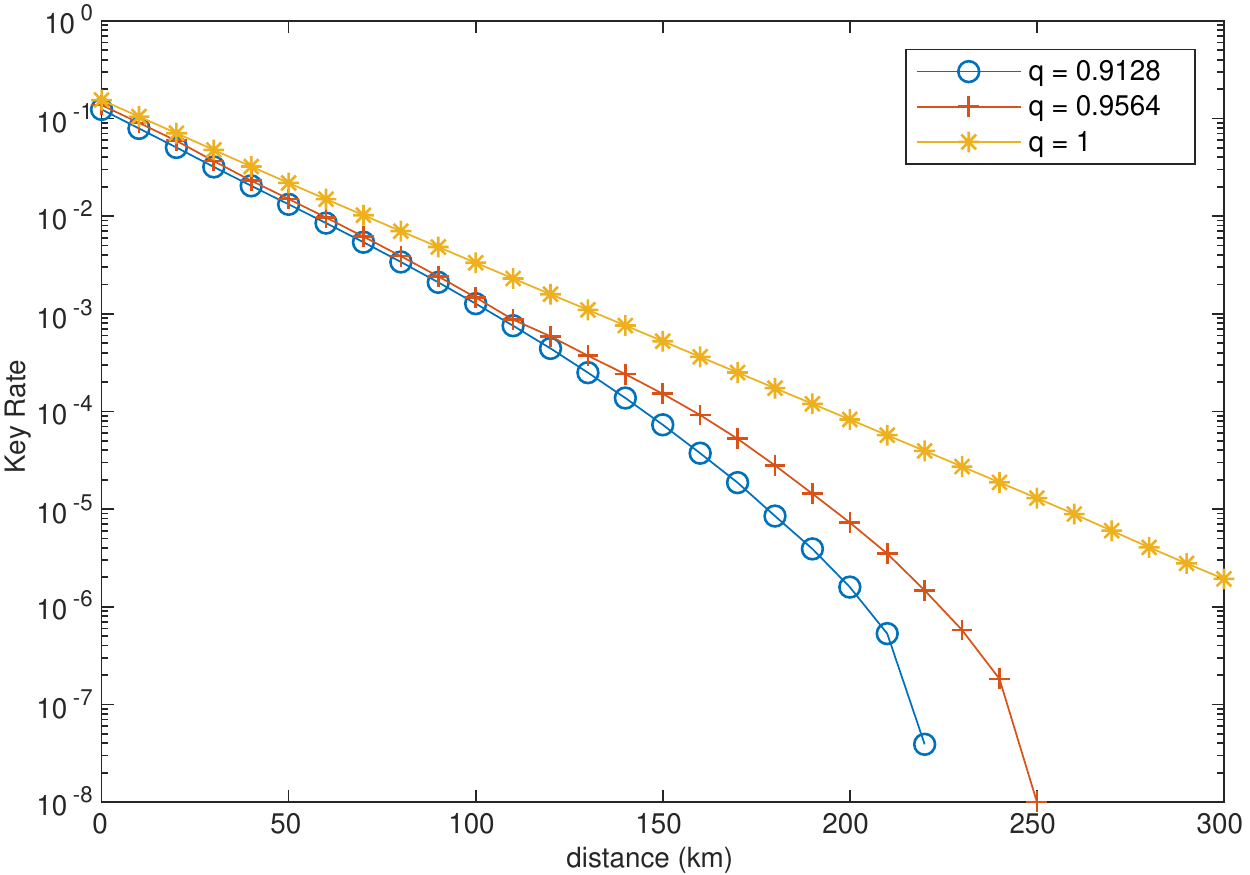}
    \caption{Comparison of the key rate of the 3 state protocol with partial and complete phase-randomisation.} \label{plots}
\end{figure}

We compared the key rate of the protocol with different degrees of partial phase-randomisation $q=0.9128$ and $q=0.9564$, and perfect phase-randomisation $q=1$ as is shown in Fig. \ref{plots}.
We observe that the key rates for the incomplete phase-randomised lasers vary significantly from the key rate for the fully phase-randomised laser. This highlights the importance of experimentally characterising the degree of phase-randomisation $q$ for high-speed QKD experiments, and using $q$ with our proof techniques to calculate the secret key rate.

The use of numerics in our methods leave some room for looseness in our results. Take for instance the bound $\epsilon_\textrm{vec}^{\mu_S\tilde{n}}$ on the closeness of the approximate eigenvectors to the true eigenvectors described in Section \ref{approxDiag}. This bound is limited by machine precision, as we choose the dimension $d$ to project onto to be large enough that $w^\mu_d$ can be upper bounded by the machine precision.
Thus, the key rate for the imperfectly phase-randomised states can be brought closer to the fully phase-randomised key rates by increasing the machine precision. Additionally, since we use the same projection while using the generalised decoy-state methods, the machine precision would also loosen the constraints in Eq. (\ref{dimReducDecoySDP}) through $\epsilon_M^\mu$.

\begin{acknowledgments}
    We thank Marcos Curty for insightful discussions regarding Ref. \cite{curty2022phasedecoy}.
    The work has been performed at the Institute for
    Quantum Computing, at the University of Waterloo,
    which is supported by Innovation, Science, and Economic
    Development Canada. The research has been supported
    by NSERC under the Discovery Grants Program, Grant
    No. 341495.
\end{acknowledgments}

\appendix

\section{Proofs related to source maps} \label{appendixSourceMaps}

In this appendix, we formally prove some results about source maps stated in the main text.

\subsection{Using source maps to lower bound the key rate} \label{appendixSourceMapKeyRateProof}

Throughout the paper, we have heavily relied on the idea of source maps. As intuitively explained in Section \ref{sectionSourceMaps}, the existence of a source map can be used to lower bound the key rate of a protocol with virtual states.

We first set up some notation before formally stating and proving the theorem. Alice's prepared state can be written as $\rho_{AA'} = \sum_{i,\mu} \ket{i,\mu}\bra{i,\mu} \otimes \rho_i^\mu$ where $A'$ is the system sent to Bob through the insecure quantum channel $\Phi$. Let $V_{\Phi}: A' \xrightarrow[]{} BE$ be the Stinespring representation of $\Phi$ so that the state shared by Alice, Bob and Eve can be written as
\begin{align}
    \rho_{ABE} = \sum_{i,\mu} \ket{i,\mu}\bra{i,\mu} \otimes \left(V_\Phi \rho_i^\mu V_\Phi^\dag\right).
\end{align}

Let $\Delta: AB \xrightarrow[]{} YZC$ be the protocol map that maps the joint state held by Alice and Bob to the raw key $Z$, Bob's measurement outcomes $Y$ along with the public announcements $C$ in that round. Thus,
\begin{align}
    \rho_{YZCE} = \left(\Delta\otimes\mathbb{I}_{E\xrightarrow[]{}E}\right)(\rho_{ABE}). \label{eqRhoYZCE}
\end{align}

The key rate can be given by the Devetak-Winter formula
\begin{align}
    R_\rho^\infty = \underset{\Phi\in \mathcal{C}_\text{real}}{\min}\  H(Z\vert CE) - \deltaleak
\end{align}
where $\mathcal{C}_\text{real}$ is the set of channels compatible with Alice and Bob's observed statistics, and $\rho_{YZCE}$ is as defined in Eq. (\ref{eqRhoYZCE}). Note that $E
$ is the auxiliary system corresponding to the Stinespring representation of the channel $\Phi$. So defining $\Phi$ specifies $E$ upto local unitaries.
We formally show that the key rate $R_\rho^\infty$ of the real protocol with signal states $\{\rho_i\}$ can be lower bounded by using a source map as stated in Section \ref{sectionSourceMaps}.
\begin{theorem}[Source map key rate]
    Define a virtual protocol with the same statistics as the real protocol, where Alice prepares the states $\{\tau_i\}$ with asymptotic key rate $R_\tau^\infty$. Let $\Psi$ be a source map connecting the virtual states to the real states such that $\rho_i = \Psi(\tau_i)$ for all $i$.
    Then $R_\tau^\infty\leq R_\rho^\infty$.
\end{theorem}
\begin{proof}
    The key rate of the virtual protocol is given by
    \begin{align}
        R_\tau^\infty = \underset{\Phi_\text{vir}\in \mathcal{C_\text{vir}}}{\min}\  H(Z\vert CE_\text{vir}) - \deltaleak
    \end{align}
    where
    \begin{align}
        \nonumber\rho_{YZCE_\text{vir}} =& \left(\Delta\otimes\mathbb{I}_{E_\text{vir}\xrightarrow[]{}E_\text{vir}}\right)\\
        &\left(\sum_{i,\mu} \ket{i,\mu}\bra{i,\mu}\otimes \left(V_{\Phi_{_\text{vir}}}\tau_i^\mu V_{\Phi_{_\text{vir}}}\right)\right)
    \end{align}
    and $\mathcal{C}_\text{vir}$ is the set of channels compatible with Alice and Bob's observed statistics.
    Let $\mathcal{C}_{\Psi} \coloneqq \{ \Phi_\text{vir} \vert \Phi_\text{vir} = \Phi \circ \Psi,\  \Phi \in \mathcal{C}_\text{real}\}$.
 
    Since the virtual protocol has the same statistics as the real protocol, $\mathcal{C}_\Psi$ is compatible with Alice and Bob's observed statistics. As a result $\mathcal{C}_\Psi \subset \mathcal{C}_\text{vir}$. 
    Additionally, it is straightforward to see that for every $\Phi\in\mathcal{C}_\text{real}$, there exists a
    $\Phi_\text{vir}\in\mathcal{C}_\Psi$.
    Thus, minimising over a smaller set of channels, can only increase the optimal value,
    \begin{align}
        \nonumber R_\tau^\infty &\leq \underset{\Phi_\text{vir}\in \mathcal{C}_\Psi}{\min}\  H(Z\vert CE_\text{vir}) - \deltaleak \\
        &= \underset{\Phi\in \mathcal{C}}{\min}\  H(Z\vert CE_\text{vir}) - \deltaleak \label{eqVirtualKeyRatePhi}
    \end{align}
    where Eq. (\ref{eqVirtualKeyRatePhi}) follows from the identification between $\mathcal{C}_\text{real}$ and $\mathcal{C}_\Psi$ made above.
    
    Let $V_\Psi: A'' \xrightarrow[]{} A'E_\text{sou}$ be the Stinespring representation of the source map $\Psi$ such that
    \begin{align}
        \rho_{ABEE_\text{sou}} = \sum_{i,\mu} \ket{i,\mu}\bra{i,\mu} \otimes \left(V_\Phi V_\Psi \tau_i^\mu V_\Psi^\dag V_\Phi^\dag\right) \label{eqRhoABEE'}
    \end{align}
    where $E_\text{vir}$ has been explicitly broken up into the individual auxiliary systems $E$ and $E_\text{sou}$ of $\Phi$ and $\Psi$ respectively.
    Thus, it follows that $\rho_{ABE} = \tr{\rho_{ABEE_\text{sou}}}{E_\text{sou}}$.
    
    The protocol map $\Delta$ is identical for both the virtual and real protocols. Additionally, the output $\rho_{YZC}$ of the protocol map is also identical since the statistics for both protocols are identical.
    Thus, $\rho_{YZCE} = \tr{\rho_{YZCEE_\text{sou}}}{E_\text{sou}}$ for all $\rho_{YZCEE_\text{sou}}$ corresponding to channels $\Phi_\text{vir}\in \mathcal{C}_{\Psi}$.
    The strong subadditivity of the conditional von Neumann entropies \cite{lieb1973proof} combined with Eq. (\ref{eqVirtualKeyRatePhi}) gives us the required inequality
    \begin{align}
        \nonumber R_\tau^{\infty} &\leq \underset{\Phi_\text{vir}\in \mathcal{C}_\Psi}{\min}\  H(Z\vert CEE_\text{sou}) - H(Z\vert Y)\\
        \nonumber &\leq \underset{\Phi_\text{vir}\in \mathcal{C}_\Psi}{\min}\  H(Z\vert CE) - H(Z\vert Y)\\
        & = \underset{\Phi\in \mathcal{C}}{\min}\  H(Z\vert CE) - H(Z\vert Y) = R_\rho^{\infty}.
    \end{align}
\end{proof}

\subsection{Construction of physical map connecting model laser state to actual laser state} \label{appendixSourceMapForLaser}

\input{SourceMapForLaser}
Note that the independence condition was important for this channel to reproduce the actual laser state. Specifically, Eq. (\ref{eqProbIndependenceUse}) would not hold for a correlated probability distribution. Thus, this technique cannot directly be used to reduce phase correlated laser states to iid states. The reduction from phase correlated laser states to an independent laser state has been done in \cite{curty2022phasedecoy}, and we thank the authors for pointing out this limitation in our methods.

\section{Bounds on projected operators}
\label{ProjBounds}

In this appendix we give the derivations for various results on projected operators that we use in Section \ref{secFinProj}.

\subsection{Bounds on one-norm} \label{projDist}

In this appendix we derive some useful bounds that will be used to prove the results in the rest of Appendix \ref{ProjBounds}. To set up notation, let $\rho$ be a density matrix, $\Pi$ be a projection with orthogonal complement $\overline{\Pi} \coloneqq \mathbb{I}-\Pi$, and $\rho^\Pi = \Pi\rho\Pi$.
Using Eq. (59) and Eq. (60) in the proof of Lemma 5 from \cite{ogawa2002new}, we get
$\norm{\Pi\rho\overline{\Pi}}_1 \leq \sqrt{1-\Tr{\rho^\Pi}}$. For notational convenience, let $W \coloneqq 1-\Tr{\rho^\Pi}$ so that we can write 
\begin{align} \label{looseOneNorm}
    \norm{\Pi\rho\overline{\Pi}}_1 \leq \sqrt{W}.
\end{align}

Note that the bound in Eq. (\ref{looseOneNorm}) depends only on $\Tr{\rho^\Pi}$. Borrowing intuition from Lemma 4 of \cite{upadhyaya2021tools}, we would expect this bound to be tighter when the state is closer to block-diagonal. Thus, we tighten the bound as follows.
\begin{theorem} \label{thmTightOneNormHBound}
    Let $\rho$ be a density matrix. With respect to a projection $\Pi$, write $\rho$ as a block-diagonal matrix
    \begin{align}
        \rho = \begin{pmatrix}
                    \Pi\rho\Pi & \Pi\rho\overline{\Pi}\\
                    \overline{\Pi}\rho\Pi & \overline{\Pi}\rho\overline{\Pi}
                \end{pmatrix}
             = \begin{pmatrix}
                    A & B\\
                    B^\dag & D
                \end{pmatrix}.
    \end{align}
    Then
    \begin{align}
        \norm{B}_1 \leq \sqrt{W} \norm{\sqrt{A}^g B \sqrt{D}^g}_\infty
    \end{align}
    where $(\cdot)^g$ denotes the generalised inverse, and $W = \Tr{\rho^{\overline{\Pi}}}$ as defined above.
\end{theorem}
\begin{proof}
    First, we briefly prove the standard result $\Pi A \Pi \geq 0$ for any $A \geq 0$ and projection $\Pi$. Equivalently, we show that $\bra{v}\Pi A\Pi\ket{v} \geq 0 \, \forall v$.
    Let $\ket{w} = \Pi \ket{v}$. Thus, $\bra{v}\Pi A\Pi\ket{v} = \bra{w}A\ket{w} \geq 0$ showing that
    \begin{align} \label{psdProj}
        \Pi A \Pi \geq 0.
    \end{align}
    In particular, this implies that $\rho^{\overline{\Pi}}\geq 0$. Since the one-norm of a positive semidefinite operator is its trace,
    \begin{align}
        \norm{\rho^{\overline{\Pi}}}_1 = \Tr{\rho^{\overline{\Pi}}} = W.
    \end{align}
    
    Define
    \begin{align*}
        \rho_\lambda
             = \begin{pmatrix}
                    A & \frac{1}{\lambda}B\\
                    \frac{1}{\lambda}B^\dag & D
                \end{pmatrix}
    \end{align*}
    where $\lambda = \norm{\sqrt{A}^g B \sqrt{D}^g}_\infty$.  We have defined $\rho_\lambda$ such that $\Tr{\rho_\lambda^{\overline{\Pi}}} = W$. Using Theorem IX.5.9 from \cite{bhatia2013matrix} gives us that $\rho_\lambda \geq 0$. Thus, we can apply the observation of Eq. (\ref{looseOneNorm}) on $\rho_\lambda$ to get
    \begin{align}
        \norm{\frac{1}{\lambda}B}_1 &\leq \sqrt{W}\\
        \implies \norm{B}_1 &\leq \lambda\sqrt{W}
    \end{align}
    which completes the proof.
\end{proof}

\begin{corollary} \label{corOffBlockDiagNorm}
    Define $H = \begin{pmatrix}
                0 & B\\
                B^\dag & 0
            \end{pmatrix}$
    as the block off-diagonal part of $\rho$. Then
    $\norm{H}_1\leq 2\sqrt{W} \norm{\sqrt{A}^g B \sqrt{D}^g}_\infty$.
\end{corollary}
\begin{proof}
    \begin{align}
        \norm{H}_1 &= \Tr{\sqrt{H^\dag H}}\\
        &= \Tr{\begin{pmatrix}
                    \sqrt{BB^\dag} & 0\\
                    0 & \sqrt{B^\dag B}
                \end{pmatrix}}\\
        &= 2\norm{B}_1\\
        &\leq 2\lambda\sqrt{W}\label{eqNormH}
    \end{align}
    where $\lambda = \norm{\sqrt{A}^g B \sqrt{D}^g}_\infty$ as defined in Theorem \ref{thmTightOneNormHBound}.
\end{proof}


\subsection{Bounds on expectation value} \label{traceBound}

Let $\Tr{P^{\overline{\Pi}}} \leq W$. Given a POVM element $\Gamma$, the proofs of Lemma \ref{lemmaExpValConstStateProj} and Lemma \ref{lemmaSMN} require upper and lower bounds of the form $\Tr{A^\Pi \Gamma}$. We derive these bounds in this appendix.

First note that $\norm{H}_1 \leq 2\lambda\sqrt{W}$ as shown in Eq. (\ref{eqNormH}). As $H$ is Hermitian, $H = H_+ - H_-$ for some $H_+, H_- \geq 0$. Since $H$ is traceless, $\Tr{H_+}=\Tr{H_-}$. Thus, 
\begin{align}
    \norm{H}_1 &= \Tr{H_+} + \Tr{H_-}\\
    &= 2\Tr{H_+}\\
    &= 2\Tr{H_-}.
\end{align}

We can then calculate the upper bound
\begin{align}
    \nonumber\Tr{A^\Pi \Gamma} =& \Tr{A\Gamma} - \Tr{A^{\overline{\Pi}}\Gamma}\\
    &- \Tr{H_+\Gamma} + \Tr{H_-\Gamma}\\
    \leq& \Tr{A\Gamma} + \Tr{H_-}\\
    =& \Tr{A\Gamma} + \lambda\sqrt{W}
\end{align}
where the inequality follows from matrix H\"older's inequality $\Tr{H_- \Gamma}\leq \norm{\Gamma}_\infty\norm{H_-}_1$, and noting that $A^{\overline{\Pi}}\geq0$ as shown in Eq. (\ref{psdProj}).
Similarly, we can compute the lower bound
\begin{align}
    \nonumber\Tr{A^\Pi \Gamma} =& \Tr{A\Gamma} - \Tr{A^{\overline{\Pi}}\Gamma}\\
    &- \Tr{H_+\Gamma} + \Tr{H_-\Gamma}\\
    \geq& \Tr{A\Gamma} - \Tr{A^{\overline{\Pi}}\Gamma} - \Tr{H_-}\\
    =& \Tr{A\Gamma} -W - \lambda\sqrt{W}
\end{align}

\subsection{Proof of Lemma \ref{lemmaExpValConstStateProj}} \label{appendixLemmaExpValConstProof}

Here we prove Lemma \ref{lemmaExpValConstStateProj} from the main text.
\lemmaExpValConstProj*
\begin{proof}
    Consider any $J\in S_\infty$ where $\Tr{({\rho_k^{\mu}}^T\otimes \Gamma_l) J} = \gamma_{l\vert k,\mu}$. From Eq. (\ref{PosConst}) we get that $(\Pi_M\otimes\eyeMeas)\;J\;(\Pi_M\otimes\eyeMeas)\geq 0$, and Lemma \ref{LemmaPartialTrace} implies that $\tr{(\Pi_M\otimes\eyeMeas)\;J\;(\Pi_M\otimes\eyeMeas)}{\mathcal{K}}\leq \Pi_M$.
    
    We now show that $(\Pi_M\otimes\eyeMeas)\;J\;(\Pi_M\otimes\eyeMeas) \in \textrm{E}_M$ by bounding its expectation values. First, note that
    \begin{align}
        \nonumber &\Tr{({\rho_k^{\mu}}^T\otimes \Gamma_l) (\Pi_M\otimes\eyeMeas)\;J\;(\Pi_M\otimes\eyeMeas)}\\
        =& \Tr{({\rho_k^{\mu M}}^T\otimes \Gamma_l) J} \label{eqStateSpaceProjJtoRhoProj}\\
        =& \Tr{\Gamma_l \Phi({\rho_k^{\mu M}}^T)} \label{eqJtoPhiChoi}\\
        =& \Tr{\Phi^\dag(\Gamma_l) {\rho_k^{\mu M}}^T} \label{eqPhiAdjoint}
    \end{align}
    where we have used the cyclic property of trace to get Eq. (\ref{eqStateSpaceProjJtoRhoProj}), and the fact that $J\in S_\infty$ is the Choi isomorphism of a channel $\Phi$ to get Eq. (\ref{eqJtoPhiChoi}).
    
    Note that for any POVM element $\Gamma_l \leq \eyeMeas$, it is the case that $\Phi^\dag (\Gamma_l) \leq \eyeState$ is also a POVM element on $\mathcal{H}$. This implies that $\norm{\Phi^\dag(\Gamma_l)}_\infty \leq 1$. Thus, we can use the results proved in Appendix \ref{traceBound} as follows.

    \underline{Lower bound:}
    \begin{align}
        \Tr{\Phi^\dag(\Gamma_l) {\rho_k^{\mu M}}^T} &\geq \Tr{\Phi^\dag(\Gamma_l) {\rho_k^{\mu}}^T} - w_{kM}^\mu - \frac{\norm{H_k^{\mu M}}_1}{2} \label{eqFirstEVMLower}\\
        &= \gamma_{l\vert k,\mu} - w_{kM}^\mu - \frac{\norm{H_k^{\mu M}}_1}{2} \label{eqJinSinftyEVMLower}\\
        &\geq \gamma_{l\vert k,\mu} - w_{kM}^\mu - \epsilon_{kM}^\mu \label{eqHnormEVMLower}
    \end{align}
    \underline{Upper bound:}
    \begin{align}
        \Tr{\Phi^\dag(\Gamma_l) {\rho_k^{\mu M}}^T} &\leq \Tr{\Phi^\dag(\Gamma_l) {\rho_k^{\mu}}^T} + \frac{\norm{H_k^{\mu M}}_1}{2} \label{eqFirstEVMUpper}\\
        &= \gamma_{l\vert k,\mu} + \frac{\norm{H_k^{\mu M}}_1}{2} \label{eqJinSinftyEVMUpper}\\
        &\leq \gamma_{l\vert k,\mu} + \epsilon_{kM}^\mu \label{eqHnormEVMUpper}
    \end{align}
    where Eq. (\ref{eqJinSinftyEVMLower}) and Eq. (\ref{eqJinSinftyEVMUpper}) follow from the expectation value constraint on $J\in\textrm{S}_\infty$, and Eq. (\ref{eqHnormEVMLower}) and Eq. (\ref{eqHnormEVMUpper}) follow from the fact that $\norm{H_k^{\mu M}}_1 \leq 2\epsilon_{kM}^\mu$ proved in Appendix \ref{projDist}. Thus, we have shown that $(\Pi_M\otimes\eyeMeas)\;J\;(\Pi_M\otimes\eyeMeas)\in\textrm{E}_M$ for all $J\in \textrm{S}_\infty$ completing the proof.
\end{proof}

\subsection{Proof of Lemma \ref{lemmaSMN}} \label{appendixLemmaSMN}

\lemmaSMN*
\begin{proof}
    Consider some $J^M\in\textrm{E}_M$. We observe that $(\eyeState\otimes\Pi_N)\;J^M\;(\eyeState\otimes\Pi_N) \geq 0$ from Eq. (\ref{PosConst}) and $\tr{(\eyeState\otimes\Pi_N)\;J^M\;(\eyeState\otimes\Pi_N)}{\mathcal{K}} \leq \Pi_M$ from Lemma \ref{LemmaPartialTrace}.
    Next we prove that $(\eyeState\otimes\Pi_N)\;J^M\;(\eyeState\otimes\Pi_N)$ satisfies the lower bound of the expectation value constraint described in Eq. (\ref{defSMN}).
    
    First, we state some preliminary results. Using Lemma \ref{lemmaExpValConstStateProj} with Eq. (\ref{eqWtOutsideJSinfty}), we show that
    \begin{align}
        \Tr{({\rho_k^{\mu}}^T\otimes \Pi_N) J^M} \geq 1- W_{kN}^\mu -w_{kM}^\mu-\epsilon^\mu_{kM} \label{projectedCrossclicks}.
    \end{align}
    Since $[\Pi_N, \Gamma_l] = 0$, we can write
    \begin{align} \label{eqCommutingGammaDecomp}
        \Gamma_l = \Gamma_l^N + \Gamma_l^{\overline{N}}
    \end{align}
    where $\Gamma_l^{\overline{N}} \coloneqq \overline{\Pi}_N \Gamma_l \overline{\Pi}_N$. We can also show that
    \begin{align}
        \Tr{\left({\rho_k^{\mu M}}^T\otimes\eyeMeas\right)J^M} &= \Tr{{\rho_k^{\mu M}}^T\tr{J^M}{\mathcal{K}}}\\
        &\leq \Tr{{\rho_k^{\mu M}}^T\Pi_M}\\
        &= \Tr{{\rho_k^{\mu M}}^T}\\
        &= 1-w_{kM}^\mu \label{eqWtOutsideProjStateJEVM}
    \end{align}
    where the inequality follows from the definition of $\textrm{E}_M$ in Lemma \ref{lemmaExpValConstStateProj}.
    Additionally, note that 
    \begin{align} \label{eqJMprojStateSame}
        \Tr{\left({\rho_k^{\mu }}^T\otimes\Gamma_l\right)J^M} = \Tr{\left({\rho_k^{\mu M}}^T\otimes\Gamma_l\right)J^M}
    \end{align}
    since $J^M\in\mathcal{B}(\mathcal{H}_M\otimes \mathcal{K})$.
    
    We can use this to estimate the lower bound as shown below.
    \begin{align}
        \nonumber &\Tr{\left({\rho_k^{\mu }}^T\otimes\Gamma_l\right)(\eyeState\otimes\Pi_N)\;J^M\;(\eyeState\otimes\Pi_N)}\\
        \nonumber=&\Tr{\left({\rho_k^{\mu}}^T\otimes\Gamma_l\right)J^M}-\Tr{\left({\rho_k^{\mu M}}^T\otimes\Gamma_l\right)J^M}\\
        &+\Tr{\left({\rho_k^{\mu M}}^T\otimes\Gamma_l^N\right)J^M} \label{eqJMprojStateSameinLowerJMN}\\
        =&\Tr{\left({\rho_k^{\mu}}^T\otimes\Gamma_l\right)J^M}-\Tr{\left({\rho_k^{\mu M}}^T\otimes\Gamma_l^{\overline{N}}\right)J^M} \label{eqCommutingGammaDecompinLowerJMN}\\
        \geq& \Tr{\left({\rho_k^{\mu}}^T\otimes\Gamma_l\right)J^M}-\Tr{\left({\rho_k^{\mu M}}^T\otimes\overline{\Pi}_N\right)J^M} \label{eqPOVMinLowerJMN}\\
        \nonumber =& \Tr{\left({\rho_k^{\mu}}^T\otimes\Gamma_l\right)J^M}-\Tr{\left({\rho_k^{\mu M}}^T\otimes\eyeMeas\right)J^M}\\
        &+ \Tr{\left({\rho_k^{\mu M}}^T\otimes\Pi_N\right)J^M}\\
        \nonumber \geq& \Tr{\left({\rho_k^{\mu}}^T\otimes\Gamma_l\right)J^M}\\
        & - (1-w_{kM}^\mu)+(1-W_{kN}^\mu-w_{kM}^\mu-\epsilon_{kM}^\mu) \label{eqWtOutsideProjStateinLowerJMN}\\
        \geq& \gamma_{l\vert k,\mu}-w_{kM}^\mu-2\epsilon^\mu_{kM}-W_{kN}^\mu \label{eqResultLowerJMN}
    \end{align}
    where Eq. (\ref{eqJMprojStateSameinLowerJMN}) follows from Eq. (\ref{eqJMprojStateSame}), Eq. (\ref{eqCommutingGammaDecompinLowerJMN}) follows from Eq. (\ref{eqCommutingGammaDecomp}), Eq. (\ref{eqPOVMinLowerJMN}) follows from the fact that $\Gamma_l\leq \mathbb{I}_\mathcal{K}$, Eq. (\ref{eqWtOutsideProjStateinLowerJMN}) follows from Eq. (\ref{projectedCrossclicks}) and Eq. (\ref{eqWtOutsideProjStateJEVM}), and Eq. (\ref{eqResultLowerJMN}) follows from the fact that $J^M\in\textrm{E}_M$.
    
    Finally, the fact that the projector commutes with the measurements $[\Pi_N,\Gamma_l] = 0$ immediately gives the upper bound
    \begin{align}
        \Tr{\left({\rho_k^{\mu}}^T\otimes\Gamma_l^N\right)J^M}&\leq \Tr{\left({\rho_k^{\mu}}^T\otimes\Gamma_l\right)J^M}\\
        &\leq \gamma_{l\vert k,\mu}+\epsilon^\mu_{kM} \label{intermediateUpper}
    \end{align}
    since $\left({\rho_k^{\mu}}^T\otimes\Gamma_l\right)-\left({\rho_k^{\mu}}^T\otimes\Gamma_l^N\right)\geq0$.
    Thus, we have shown that for all $J^M\in\textrm{EV}_M$, $(\eyeState\otimes\Pi_N)\;J^M\;(\eyeState\otimes\Pi_N)\in\textrm{S}_{MN}$ completing the proof.
\end{proof}

\subsection{Closeness of eigenvectors} \label{appendixEigenBound}

As in this paper, one might run into a situation where diagonalising a density matrix $\rho$ is of interest, while a perturbed density matrix $\sigma = \rho+H$ can be diagonalised where $\norm{H}_1\leq2\epsilon$. In this appendix we explain how and when one can approximate the eigenvectors of $\rho$ with the eigenvectors of $\sigma$.

Let $\lambda_i(S)$ be the $i^\textrm{th}$ largest eigenvalue of a compact, self-adjoint operator $S$. From Theorem 4.10 of \cite{teschl2009mathematical}, we can write the eigenvalues as
$$\lambda_n(S) = \min_{\{\ket{\psi_1},\ldots,\ket{\psi_{n-1}}\}}\quad \max_{\ket{\psi}\in P^\perp(\ket{\psi_1},\ldots,\ket{\psi_{n-1}})}\bra{\psi}S\ket{\psi}$$
where $P^\perp(\ket{\psi_1},\ldots,\ket{\psi_{n-1}}) \coloneqq \{\ket{\psi}\in\textrm{span}\{ \ket{\psi_1},\ldots,\ket{\psi_{n-1}}\}^\perp\vert \norm{\psi} = 1\}$ is the space perpendicular to the vectors $\ket{\psi_1},\ldots,\ket{\psi_{n-1}}$. From this we can bound the change in eigenvalues due to the perturbation.
\begin{theorem}\label{Weyl}
    Let $\mathcal{H}$ be a Hilbert space. Given $\rho\in\textrm{D}(\mathcal{H})$, $\sigma\in\textrm{D}(\mathcal{H}$) and $H = \sigma-\rho$ with $\norm{H}_1\leq 2\epsilon$ as defined above,
    $$\abs{\lambda_i(\rho)-\lambda_i(\sigma)} \leq \epsilon$$
    for all eigenvalues indexed by $i$.
\end{theorem}
\begin{proof}
    The proof follows similarly to the proof of Weyl's inequality, which is for finite dimensions.
    \begin{align}
        \nonumber \lambda_i(\sigma) &= \min_{\{\ket{\psi_1},\ldots,\ket{\psi_{i-1}}\}} \max_{\ket{\psi}\in P^\perp(\ket{\psi_1},\ldots,\ket{\psi_{i-1}})}\bra{\psi}S\ket{\psi}\bra{\psi}\sigma\ket{\psi}\\
        \nonumber &= \min_{\{\ket{\psi_1},\ldots,\ket{\psi_{i-1}}\}} \max_{\ket{\psi}\in P^\perp(\ket{\psi_1},\ldots,\ket{\psi_{i-1}})}\\
        \nonumber&\phantom{spaceasdspacespacespacespace}\left(\bra{\psi}\rho\ket{\psi}+\bra{\psi}H\ket{\psi}\right)\\
        \nonumber &\leq \min_{\{\ket{\psi_1},\ldots,\ket{\psi_{i-1}}\}}\\
        \nonumber &\phantom{\min_{\{\ket{\psi_1},\ldots}}\left(\max_{\ket{\psi}\in P^\perp(\ket{\psi_1},\ldots,\ket{\psi_{i-1}})}\bra{\psi}\rho\ket{\psi}+\norm{H}_\infty\right)\\
        \nonumber &= \lambda_i(\rho)+\norm{H}_\infty\\
        &\leq \lambda_i(\rho) +\frac{\norm{H}_1}{2} \label{loosenessOneNorm}\\
        &= \lambda_i(\rho) + \epsilon
    \end{align}
    where Eq. (\ref{loosenessOneNorm}) follows from noting that $\Tr{H} = 0$.
    Starting with $\rho$ instead of $\sigma$ in the first line and following the same steps while replacing $H$ with $-H$ gives us $\lambda_i(\rho) \leq \lambda_i(\rho)+\epsilon$. Combining both together, we get $\abs{\lambda_i(\rho)-\lambda_i(\sigma)}\leq \epsilon$ as stated.
\end{proof}

Before talking about the individual eigenvectors, we shall introduce the Davis-Kahan theorem \cite{DavisKahan}. The intuition of the theorem can be understood as follows. Any density operator can be diagonalised as
$$\tau = W_{\tau} D_{\tau} W^\dag_{\tau}$$ where $W_{\tau}$ is a unitary whose columns are the eigenvectors of $\tau$, and $D_\tau$ is a diagonal operator whose elements are eigenvalues of $\tau$.
The unitary can be written as a block matrix
$$W_{\tau} = [W\quad W_\perp]$$
where $W$ and $W_\perp$ are isometries whose columns span eigenspaces of $\tau$. If $\tau$ is not degenerate, these eigenspaces are orthogonal to each other. The density operator $\tau$ can be written as
$$\tau = W\tau_0 W^\dag + W_\perp\tau_1 W_\perp^\dag$$
where $\tau_0$ and $\tau_1$ are diagonal matrices whose elements are the eigenvalues of $\tau$ corresponding to the eigenvectors in $W$ and $W_\perp$ respectively.

Let $\rho$ and $\sigma$ have decompositions with $U_\rho = [U\quad U_\perp]$ and $V_\sigma =[V\quad V_\perp]$. The theorem then formalises the intuition that if $\rho$ and $\sigma$ are ``close", then the eigenspaces spanned by $U$ and $V_\perp$ are ``almost" orthogonal.

\begin{theorem}[Davis-Kahan] \label{DK}
Let $\rho =U\rho_0U^\dag+U_\perp\rho_1U^\dag_\perp$ and $\sigma = V\sigma_0V^\dag+V_\perp\sigma_1V_\perp^\dag$ be density operators where the block matrices $[U\quad U_\perp]$ and $[V\quad V_\perp]$ are unitaries. Let $H = \sigma - \rho$. If the eigenvalues of $\rho_0$ are contained in an interval $(a,b)$, and the eigenvalues of $\sigma_1$ are excluded from the interval $(a-\delta,b+\delta)$ for some $\delta>0$, then
\begin{align}
    \norm{V_\perp^\dag U}\leq \frac{\norm{V_\perp^\dag H U}}{\delta}
\end{align}
for any unitarily invariant norm $\norm{\cdot}$.
\end{theorem}

Although the proof in \cite{DavisKahan} is for finite dimensions, the proof for infinite-dimensional density operators is exactly the same.
Intuitively, the $\delta$ represents how separated the eigenspaces of $\sigma$ are relative to the perturbation $\epsilon$. If this $\delta$ is too small, the corresponding eigenspaces of $\rho$ and $\sigma$ could be quite different.
Instructive examples and further intuition about this theorem can be found in \cite{DavisKahan}.

\begin{corollary} \label{corFidelityEigenvectors}
    Let $\rho$, $\sigma$ be density operators with $H = \sigma - \rho$ and $\norm{H}_1 \leq 2 \epsilon$. Define $\delta_i = \min \{\lambda_i(\sigma)-\lambda_{i-1}(\sigma) - \epsilon,\ \lambda_{i+1}(\sigma) - \lambda_i(\sigma)-\epsilon\}$. Then
    \begin{align}
        F(U_iU_i^\dag,V_iV_i^\dag) \geq 1- \frac{\epsilon^2}{\delta_i^2}
    \end{align}
    where $U_i$ and $V_i$ are the $i^\textrm{th}$ eigenvectors of $\rho$ and $\sigma$ respectively.
\end{corollary}
\begin{proof}
    For each $i$, let $\rho$ and $\sigma$ have decomposition
    \begin{align}
        \rho &= U_i \lambda_i(\rho)U_i^\dag+{U_i}_\perp \rho_1 {{U_i}_\perp^\dag}\\
        \sigma &= V_i \lambda_i(\sigma)V_i^\dag+{V_i}_\perp \sigma_1 {{V_i}_\perp^\dag}
    \end{align}
    as described in Theorem \ref{DK}.
    A direct consequence of Theorem \ref{Weyl} is that $\lambda_i(\rho)$ lies in the interval $(a_i,b_i)$ with $a_i = \lambda_i(\sigma)-\epsilon$ and $b_i =\lambda_i(\sigma)+\epsilon$. Additionally, it can be easily verified that all the eigenvalues of $\sigma_1$ lie outside the interval $(a_i-\delta_i, b_i + \delta_i)$.
    Thus, using Theorem \ref{DK}
    \begin{align}
        \norm{{{V_i}_\perp^\dag} U_i}_\infty &\leq \frac{\norm{{{V_i}_\perp^\dag} HU_i}_\infty}{\delta_i}\\
        &\leq \frac{\norm{{{V_i}_\perp^\dag}}_\infty \norm{H}_\infty \norm{U_i}_\infty}{\delta_i}\\
        &= \frac{\norm{H}_\infty}{\delta_i}\\
        &\leq \frac{\norm{H}_1}{2\delta_i}\\
        &\leq\frac{\epsilon}{\delta_i} \label{DKbound}
    \end{align}
    where the second inequality follows from the fact that the $\infty$-norm is submultiplicative $\norm{AB}_\infty\leq \norm{A}_\infty \norm{B}_\infty$, and the succeeding equality is a consequence of the fact that $\norm{W}_\infty = 1$ for any isometry $W$.
    
    Now consider the diagonalising unitaries
    $U_{\rho} =[U_i\quad {U_{i}}_{\perp}]$ and $V_{\sigma} =[V_i\quad {V_{i}}_{\perp}]$. Thus,
    $$W \coloneqq U_\rho^\dag V_\sigma = 
    \begin{pmatrix}
        U_i^\dag V_i& U_i^\dag {V_i}_\perp \\
        {U_i}_\perp^\dag V_i & {U_i}_\perp^\dag {V_i}_\perp
    \end{pmatrix}$$ must also be unitary. So $W W^\dag = \mathbb{I}$. Looking at the first block of $WW^\dag$ which is 1-dimensional,
    \begin{align} \label{block}
        U_i^\dag V_i V_i^\dag U_i+U_i^\dag {V_i}_\perp {V_i}_\perp^\dag U_i = 1.
    \end{align}
    
    Thus,
    \begin{align}
        \abs{1-U_i^\dag V_i V_i^\dag U_i} &= \norm{1-U_i^\dag V_i V_i^\dag U_i}_\infty \label{eqAbsInfNorm}\\
        &= \norm{U_i^\dag {V_i}_\perp {V_i}_\perp^\dag U_i}_\infty\\
        &\leq \norm{U_i^\dag {V_i}_\perp}_\infty \norm{{V_i}_\perp^\dag U_i}_\infty\\
        &\leq \frac{\epsilon^2}{\delta_i^2} \label{overlap}
    \end{align}
    where Eq. (\ref{overlap}) follows from Eq, (\ref{DKbound}).
    Observe that the fidelity between the eigenvectors $U_i$ and $V_i$ is $\abs{U_i^\dag V_i}$. Eq. (\ref{overlap}) then directly gives us a bound on the fidelity,
    \begin{align}
        F(U_iU_i^\dag,V_iV_i^\dag)^2 \geq 1-\frac{\epsilon^2}{\delta_i^2}.
    \end{align}
\end{proof}

Theorem \ref{thmApproxDiag} from the main text is just a special case of Theorem \ref{Weyl}, and Corollary \ref{corFidelityEigenvectors} as follows. Let $\sigma = \rho^\Pi + \rho^{\overline{\Pi}}$ and $H = \sigma-\rho$. Corollary \ref{corOffBlockDiagNorm} gives $\norm{H}_1 \leq 2 \norm{\sqrt{\rho^\Pi}^g \Pi\rho\overline{\Pi} \sqrt{\rho^{\overline{\Pi}}}^g}_\infty \sqrt{\Tr{\rho^{\overline{\Pi}}}}$. Thus, defining $\epsilon_\textrm{proj} = \norm{\sqrt{\rho^\Pi}^g \Pi\rho\overline{\Pi} \sqrt{\rho^{\overline{\Pi}}}^g}_\infty \sqrt{\Tr{\rho^{\overline{\Pi}}}}$, Theorem \ref{thmApproxDiag} is exactly Theorem \ref{Weyl} and Corollary \ref{corFidelityEigenvectors}.

As a final remark, we note that all the results in this appendix hold if we replace $\epsilon$ with $\epsilon_\infty$ where $\norm{H}_\infty\leq \epsilon_\infty$.

\section{Bound on weight outside projected subspace} \label{appendixBoundW}

The general method of using ``cross-clicks" to bound the weight outside the projected subspace is taken from Chapter 2 of \cite{narasimhachar2011study}.

Since we use threshold detectors, Bob's measurements are block-diagonal in the total photon number of the two pulses.
Given the probability that Bob received a state with $n$ photons $p(n)$, the probability $p(\textrm{event})$ of observing a particular detection event can then be written as
\begin{align}
    p(\textrm{event}) =& \sum_{n=0}^\infty p(n) p(\textrm{event}\vert n)\\
    \nonumber =& \sum_{n=0}^N p(n) p(\textrm{event}\vert n)\\
    &+ \sum_{n=N+1}^\infty p(n) p(\textrm{event}\vert n)\\
    \nonumber \geq&\  p(\leq N) p^\textrm{min}(\textrm{event}\vert \leq N)\\
    &+ p(> N) p^\textrm{min}(\textrm{event}\vert > N)
\end{align}
where $p^\textrm{min}(\textrm{event}\vert \leq N)$ denotes the minimum probability of observing the detection event given the state has $\leq N$ photons. Using the fact that $p(\leq N)+p(>N) = 1$ and rearranging we get
\begin{align} \label{boundW}
    p(>N) \leq \frac{p(\textrm{event})-p^\textrm{min}(\textrm{event}\vert \leq N)}{p^\textrm{min}(\textrm{event}\vert > N)-p^\textrm{min}(\textrm{event}\vert \leq N)}.
\end{align}
So in order to bound the weight outside the $\leq N$ subspace, we need to find $p(\textrm{event}\vert n)$.

We have some choice when choosing the specific event, which we call a ``cross-click" event. Here, we define a cross-click to be any click pattern that records a click in both the detectors while ignoring all clicks in mode $d_2$. We make this choice because it makes the calculations simpler as shall become apparent. We do not claim that this is the optimal choice. However, as shown above, the validity of the bounds in Eq. (\ref{boundW}) are independent of the choice of detection event.
We wish to bound the probability $p(\textrm{cc}\vert n)$ of cross-clicks over all input states with $n$ total photons in both pulses. Although this task is hard for a generic choice of cross-click event, our specific choice allows us to simplify the task with the following observation.
The probability of cross-clicks $p(\textrm{cc})$ does not depend on either the phase or the relative phase of the two pulses. Thus, without loss of generality, we can always consider individually phase-randomised pulses without changing the statistics. In other words, we can assume that our input state is a probabilistic mixture of $\ket{m,n-m}\bra{m,n-m}$ where the total photon number is $n$. Thus, it is sufficient to bound the probability $p(\textrm{cc}\vert n)$ of cross-clicks over all input states with $m$ photons in the first pulse, and $n-m$ photons in the second.

\begin{figure}[h]
    \centering
    \includegraphics[width = \linewidth]{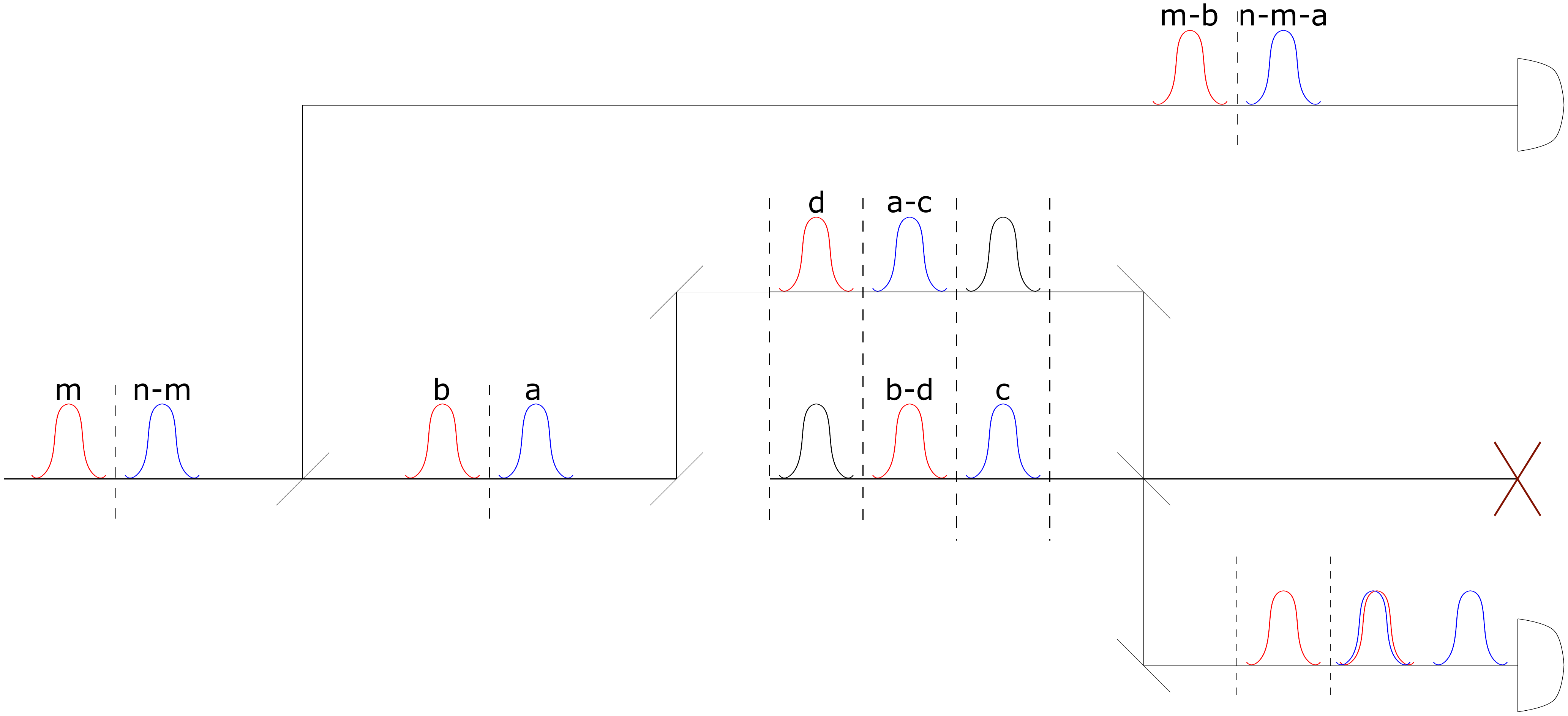}
    \caption{Bob receives $m$ and $n-m$ photons in the two pulses. Of that $a+b$ photons go into the X basis measurement line, and $n-a-b$ go to the Z basis detector. $c$ and $d$ photons go into the outside time-bins of the interferometer with the rest going into the middle time-bin.} \label{figBobMeasurementCross}
\end{figure}

As shown in Fig. \ref{figBobMeasurementCross}, the probability of a cross-click given an input state containing $m$ and $n-m$ photons in the two pulses is
\begin{align}
    \nonumber p(\textrm{cc}\vert m,n-m) =& \sum_{\substack{a+b\neq 0\\a+b\neq n}} \binom{m}{b}\binom{n-m}{a}t^{a+b}(1-t)^{n-a-b}\\
    &\left(\frac{1}{2}\right)^{a+b}\sum_{c+d\neq 0}\binom{a}{c}\binom{b}{d}\left(1-\left(\frac{1}{2}\right)^{c+d}\right).
\end{align}
Here, $\binom{m}{b}t^{b}(1-t)^{m-b}$ factor reflects the probability of $m$ input photons being split into $m-b$ and $b$ photons. The $\binom{b}{d}\left(\frac{1}{2}\right)^b$ factor reflects the probability of $b$ photons being split into $d$ and $b-d$ photons for the two arms of the interferometer. Similarly, the same reasoning applies for the input pulse with $n-d$ photons. The last $\left(1-\left(\frac{1}{2}\right)^{c+d}\right)$ factor is to subtract the case when all $c+d$ photons go into the line with the detector we do not use.

We first calculate the second summation,
\begin{align}
    S(a,b) &= \sum_{c+d\neq 0}\binom{a}{c}\binom{b}{d}\left(1-\left(\frac{1}{2}\right)^{c+d}\right)\\
    &= \sum_{c=0}^a\sum_{d=0}^b\binom{a}{c}\binom{b}{d}\left(1-\left(\frac{1}{2}\right)^{c+d}\right)-0\\
    &= 2^{a+b}-\left(\frac{3}{2}\right)^{a+b}.
\end{align}
Thus, the cross-click probability can be simplified as
\begin{align}
    \nonumber p(\textrm{cc}\vert m,n-m) =& \sum_{\substack{a+b\neq 0\\a+b\neq n}} \binom{m}{b}\binom{n-m}{a}t^{a+b}(1-t)^{n-a-b}\\
    &\left(\frac{1}{2}\right)^{a+b} S(a,b)\\
    =& \sum_{\substack{a+b\neq 0\\a+b\neq n}} \nonumber\binom{m}{b}\left(\frac{t}{2}\right)^{b}(1-t)^{m-b}\binom{n-m}{a}\\
    &\left(\frac{t}{2}\right)^{a}(1-t)^{n-m-a}\; S(a,b). \label{crossIntermediate}
\end{align}
In order to simplify this, we compute
\begin{align}
    f(x,m) &= \sum_{b=0}^m \binom{m}{b}\left(\frac{t}{2}\right)^{b}(1-t)^{m-b}x^b\\
    &= \left(1-t+\frac{xt}{2}\right)^{m}. \label{f(x,m)}
\end{align}
So using Eq. (\ref{f(x,m)}) in Eq. (\ref{crossIntermediate}) we get
\begin{align}
    \nonumber p(\textrm{cc}\vert m,n-m) =& f(2,m)f(2,n-m)\\
    \nonumber &-f\left(\frac{3}{2},m\right)f\left(\frac{3}{2},n-m\right)\\
    &-\left(\frac{2t}{2}\right)^n+\left(\frac{3t}{4}\right)^n\\
    \nonumber =& \left(1-t+\frac{2t}{2}\right)^{n}-\left(1-t+\frac{3t}{4}\right)^{n}\\
    &- t^n +\left(\frac{3t}{4}\right)^n\\
    =& 1 - t^n - \left(1-\frac{t}{4}\right)^{n}+\left(\frac{3t}{4}\right)^n.
\end{align}
We observe that the cross-click probability does not depend on $m$ which intuitively follows from the symmetry of our definition of cross-clicks.

Viewing the cross-click probability as a function of $n$
\begin{align}
    f(n) = 1 - t^n - \left(1-\frac{t}{4}\right)^{n}+\left(\frac{3t}{4}\right)^n,
\end{align}
we look to show that the function is monotonically increasing. This would make it easy to find $p^\textrm{min}(\textrm{cc}\vert \leq N)$.
We do this by considering
\begin{align}
    \nonumber f(n+1)-f(n) =& t^n(1-t)+\left(1-\frac{t}{4}\right)^n\left(\frac{t}{4}\right)\\
    &- \left(\frac{3t}{4}\right)^n\left(1-\frac{3t}{4}\right)
\end{align}
and showing that this is positive for all positive integers $n$.
We first note that $0\leq t\leq 1$ which gives us
\begin{align}
    t&\leq1\\
    1-t&\geq 0\\
    1-\frac{t}{4}-\frac{3t}{4}&\geq 0\\
    1-\frac{t}{4}&\geq \frac{3t}{4}.
\end{align}
Raising both sides to the $n^\textrm{th}$ power and multiplying both sides of the inequality by $\frac{t}{4}$,
\begin{align}
    \left(1-\frac{t}{4}\right)^n\frac{t}{4}&\geq \left(\frac{3t}{4}\right)^n\frac{t}{4}\\
    \left(1-\frac{t}{4}\right)^n\frac{t}{4}-\left(\frac{3t}{4}\right)^n\left(1-\frac{3t}{4}\right)&\geq \left(\frac{3t}{4}\right)^n(t-1).
\end{align}
Finally, adding $t^n(1-t)$ to both sides of the equation,
\begin{align}
    \nonumber &t^n(1-t)+\left(1-\frac{t}{4}\right)^n\frac{t}{4}-\left(\frac{3t}{4}\right)^n\left(1-\frac{3t}{4}\right)\\
    &\geq t^n(1-t)+\left(\frac{3t}{4}\right)^n(t-1)\\
    f(n+1)-f(n) &\geq \left(t^n-\left(\frac{3t}{4}\right)^n\right)(1-t)\geq 0
\end{align}
where the last inequality follows from the fact that $t\geq \frac{3t}{4}$.
Thus, $p^\textrm{min}(\textrm{cc}\vert \leq N) = p(\textrm{cc}\vert 0)=0$, and $p^\textrm{min}(\textrm{cc}\vert > N) = p(\textrm{cc}\vert N+1)$. Using this in Eq. (\ref{boundW}),
\begin{align}
    p(>N) \leq \frac{p(\textrm{cc})}{1-t^{N+1}-\left(1-\frac{t}{4}\right)^{N+1}+\left(\frac{3t}{4}\right)^{N+1}}
\end{align}
where we can obtain the cross-click probability $p(\textrm{cc})$ from the observations.

\bibliography{main}
\end{document}

%% file: GeneralDecoyQKDFlowChart.tex
\begin{tikzpicture}[node distance = 2cm, every node/.style={scale=0.6},font = \large]
        
        
        \node [io, align = left] (rho) {$\rho^{\mu_S} = \sum_{\tilde{n}} p_{\tilde{n}}\ket{\tilde{n}}\bra{\tilde{n}}$};
        
        \node [decision, right of=rho,xshift = 3cm] (diagDec) {Is it hard to find $p_{\tilde{n}}$ and $\ket{\tilde{n}}$?};
        
        \node [processDiag, below of=diagDec, yshift = -4cm] (approxDiag) {\underline{Approx. Diagonalise}
        $\rho' = \sum_{\tilde{n}} p'_{\tilde{n}} \ket{v_{\tilde{n}}}\bra{v_{\tilde{n}}}$
        $\abs{p'_{\tilde{n}}-p_{\tilde{n}}}\leq \epsilon_\textrm{proj}$
        $\norm{\ket{v_{\tilde{n}}}\bra{v_{\tilde{n}}}-\ket{\tilde{n}}\bra{\tilde{n}}}_1 \leq \epsilon_\textrm{vec}^{\tilde{n}}$};
        
        \node[circle, draw,right of= diagDec, xshift = 2cm](connect){};
        
        \node[process, right of = connect, xshift = 3cm](tagging){Block-tagging};
        
        \node[process, right of = tagging, xshift = 3cm, below of = tagging, yshift = -1cm](blockKey){Block key rate $R_{\tilde{n}}$};
        \node[processDiag, below of = blockKey, yshift = 0.69cm](blockKeyDiag){Use $p'_{\tilde{n}}-\epsilon_\textrm{proj}$ and $\rho_A^{v_{\tilde{n}}} + \epsilon_\textrm{vec}^{\tilde{n}} S$ since $p_{\tilde{n}}$ and $\rho_A^{\tilde{n}}$ are not known}; 
        \node[circle,radius = 0.5cm, draw=none,below of= blockKey, yshift = 1.49cm, left of = blockKey, anchor = east,xshift = -0.17cm](blockLeft){};
        \node[circle,radius = 0.5cm, draw=none,below of= blockKey, yshift = 1.49cm, right of = blockKey, anchor = west, xshift = 0.17cm](blockRight){};
        
        \node[circle, draw,left of= blockLeft, xshift = -0.66cm](connectToBlockKey){};
        
        \node[process, below of = blockKeyDiag, xshift = -5cm, yshift = -1cm](genDecoy){Generalised decoy-state method with finite projections}; 
        \node[processDiag, below of = genDecoy, yshift = 0.7cm](genDecoyDiag){Use $\ket{v_{\tilde{n}}}$ and loosen bounds by $\epsilon_\textrm{vec}^{\tilde{n}}$};
        \node[circle,radius = 0cm, draw=none,below of= genDecoy, yshift = 1.25cm, left of = genDecoy, anchor = east,xshift = -0.17cm](decoyLeft){};
        
        
        \node[io, right of = blockKey, xshift = 4cm,yshift = -0.5cm, align = left](keyRate){Key Rate\\
                            $R = \sum_{\tilde{n}}^{\tilde{N}}R_{\tilde{n}}$};
        
        \draw [arrow] (rho) -- (diagDec);
        \draw [arrowDiag] (diagDec) --node[anchor=east] {yes} (approxDiag);
        \draw [arrow] (diagDec) --node[anchor=south] {no} (connect);
        \draw [arrowDiag] (approxDiag) -| (connect);
        \draw [arrow] (connect) -- (tagging);
        \draw [arrow] (tagging.245) -- (genDecoy.108);
        
        
        \draw [arrow] (tagging) -- (connectToBlockKey);
        \draw [arrow] (genDecoy) -- (connectToBlockKey);
        \draw [arrow] (connectToBlockKey) -- (blockLeft);
        \draw [arrow] (blockRight) -- (keyRate);
    \end{tikzpicture}

%% file: SourceMapForLaser.tex
Define the following for notational convenience:
\begin{align}
    \rho_{\phi} &\coloneqq \ket{\sqrt{\mu}\exp{i \phi}}\bra{\sqrt{\mu}\exp{i \phi}}\\
    U_{\theta} &: \ket{\sqrt{\mu}\exp{i \phi}} \mapsto \ket{\sqrt{\mu}\exp{i \phi + \theta}}
\end{align}
where $U_{\theta}$ represents the action of a phase modulator which is unitary. Let $\rhoPR = \intd{} \frac{1}{2\pi} \rho_{\phi}$ be the fully phase-randomised state and $\rhomodel = q \rhoPR + (1-q) \ket{\sqrt{\mu}}\bra{\sqrt{\mu}}$.
Also define
\begin{align}
    \tau_{\phi} &\coloneqq U_\phi \rhomodel U_\phi^\dag, \mathrm{ and} \\
    \tpn &\coloneqq \tpo \ldots \tp{n}
\end{align}
where $\tp{i} \coloneqq \frac{\p{i}-q/2\pi}{1-q}$.

\begin{lemma}
    Define 
    \begin{align}
        \nonumber \Phi(\sigma^n) \coloneqq \intdn \tpn \\
        \left(\Uphi{1} \ldots \Uphi{n}\right) \sigma^n \left(\Udagphi{1} \ldots \Udagphi{n}\right).    
    \end{align}
    Then
    \begin{enumerate}
        \item If $q \leq 2\pi \; \underset{i}{\min} \;  \underset{\phi_i}{\min} \; \p{i}$, then $\Phi$ is a mixed unitary channel. \label{probcond}
        \item If the actual laser state is phase-independent across pulses, i.e. $\p{i} = p_{\Phi_i}(\phi_i)$ for all i, then $$\Phi\left(\rhomodel^{\otimes n}\right) = \rho_{\mathrm{laser}}$$ where $\rho_{\mathrm{laser}}$ is as defined in Eq. (\ref{generalLaser}). \label{mapcond}
    \end{enumerate}
\end{lemma}
\begin{proof}
    Condition \ref{probcond}.
    
    Verifying that $\tilde{p}$ is a probability density function is straightforward. This directly implies that $\Phi$ is a mixed unitary channel.
    
    Condition \ref{mapcond}.
    
    First note that $\Uphi{i} \rhoPR \Udagphi{i} = \rhoPR$ for all $\phi_i\in [0,2\pi)$ where $\rhoPR$ is the fully phase-randomised state. Using this, a straightforward computation gives
    \begin{align}
        \notag &\intd{i} \tp{i} \tauphi{i}\\
        =& \intd{i} \p{i} \rhophi{i} \label{singleIntegral}
    \end{align}
    
    Now looking at the action of the map on the model state,
    \begin{align}
        \nonumber\Phi\left(\rhomodel^{\otimes n}\right) =& \intdn\\
        \notag&\phantom{\intd{1}}\tpn \\
        &\left(\Uphi{1} \ldots \Uphi{n}\right) \rhomodel^{\otimes n} \left(\Udagphi{1} \ldots \Udagphi{n}\right)\\
        \nonumber=& \intd{1} \tpo \tauphi{1} \otimes \ldots \\
        &\otimes\intd{n} \tilde{p}_{\Phi_n}(\phi_n) \tauphi{n} \label{eqProbIndependence}\\
        \nonumber=& \intd{1} \po \rhophi{1}\otimes \ldots \\
        & \otimes\intd{n} p_{\Phi_n}(\phi_n) \rhophi{n} \label{eqProbIndependenceUse}\\
        \nonumber=& \intdn \\
        &\pn \rhophi{1}\otimes \ldots \otimes \rhophi{n}\\
        =& \rho_\mathrm{laser}
    \end{align}
    where Eq. (\ref{eqProbIndependence}) follows from the assumption that the probability distribution is independent, and Eq. (\ref{eqProbIndependenceUse}) follows from Eq. (\ref{singleIntegral}).
\end{proof}